\documentclass[12pt]{article}
\usepackage[top=1.2in, bottom=1.2in, left=1in, right=1in]{geometry}
\usepackage{graphicx}
\usepackage{booktabs}
\usepackage{subcaption}
\usepackage{comment}
\usepackage{tikz}
\usepackage[figuresright]{rotating}
\usepackage{tablefootnote}
\usepackage{diagbox}
\usepackage[sort&compress,authoryear]{natbib}
\usepackage[colorlinks,
linkcolor=black,
anchorcolor=black,
citecolor=black,
breaklinks=true
]{hyperref}
\usepackage[titletoc,title]{appendix}
\usepackage{amsmath,amsthm,amsfonts,amssymb,bm} 
\usepackage{nccmath}
\usepackage{subcaption}

\usepackage{makeidx} 

\usepackage{listings}                          
            
\usepackage{psfrag} 
\usepackage{epstopdf}
\usepackage{enumitem}

\theoremstyle{definition}
\newtheorem{thm}{Theorem}[section]

\newtheorem{prop}[thm]{Proposition}

\newtheorem{coro}[thm]{Corollary}
\newtheorem{lemma}[thm]{Lemma}

\newtheorem{Assumption}[thm]{Assumption}

\def\E{{\mathbb{E}}}
\def\P{{\mathbb{P}}}

\title{Non-concave Corporate Management with Option Incentives under Value-at-Risk Constraint}  
\author{
Wenyuan Li$^{a}$ \and
Haoqi Lyu$^{a,*}$ \and
Pengyu Wei$^{b}$
}

\date{}
\begin{document} 
\maketitle

\begin{center}
\small
$^{a}$Department of Statistics and Actuarial Science, The University of Hong Kong, Pok Fu Lam Road, Hong Kong SAR, China\\
\texttt{wylsaas@hku.hk}, \texttt{lyuhq@connect.hku.hk}\\[0.5em]
$^{b}$Division of Banking and Finance, Nanyang Business School, Nanyang Technological University, Singapore\\
\texttt{pengyu.wei@ntu.edu.sg}\\[0.5em]
$^{*}$Corresponding author \qquad 
\end{center}

\begin{abstract}
This article studies a dynamic corporate risk management problem by considering the decision-making of risk-averse managers who exert costly effort and select project risk. We study how a Value-at-Risk (VaR) constraint affects managerial decisions and the distribution of firm value when the manager’s objective is non-concave with a fixed salary and options.  By the concavification technique, we analyze the optimal terminal firm value on the concave envelope of the objective function. Applying the quantile formulation and the martingale approach, we can derive explicit solutions for optimal effort, terminal firm value, and project choice. The optimal terminal firm value can be divided into nine cases by carefully discussing the choices of VaR floor and tail probability. Compared with the benchmark case, we find that a VaR manager will smooth terminal firm value across states, reducing it in good states while supporting it in adverse states. Moreover, a VaR requirement generally improves downside protection and reduces bankruptcy probability when the VaR floor is low or moderate. However, when the VaR floor is sufficiently high, it can increase bankruptcy probability and induce gambling-for-recovery behavior in adverse states. Our sensitivity analysis indicates that greater managerial effort uniformly improves firm value. Moreover, more incentive options make managers more responsible, leading to a smoother terminal firm value across states. In contrast, a high fixed salary makes the manager less responsible and ultimately causes a more dispersed firm value.
\end{abstract}


\section{Introduction}
Managers play a central role in a firm's performance. Their decisions influence both the growth of firm value and the firm’s risk exposure. However, because managers generally maximize their own objectives rather than shareholder value, their interests may not be fully aligned with those of the firm's owners. This conflict has long been a central issue in corporate finance. \cite{JENSEN1976305} identify the separation of ownership and control as a fundamental source of agency conflict. \cite{holmstrom1979moral} further shows that when managerial actions are not directly observable, moral hazard becomes an important concern. 

Motivated by this tension, many studies have investigated managerial decisions and their effects on firm value in a dynamic setting. \cite{carpenter2000does} analyzes the optimal investment strategy of a risk-averse fund manager compensated with a call option and shows that the option compensation does not necessarily increase risk-taking; in some cases, the manager optimally chooses lower volatility than they would without the option. \cite{cadenillas2004leverage, cadenillas2007optimal} develop
a dynamic firm value model in which managers choose effort and firm risk, and study the effect of compensation design on managerial risk-taking behavior.
\cite{bensoussan2015entrepreneurial} examine entrepreneurial decision problems with non-concave objectives, and characterize the optimal choices of effort and project risk by the martingale method and concavification technique. While the existing literature provides an extensive investigation into managerial effort, compensation, and risk-taking behavior, it generally does not incorporate explicit firm-level risk management constraints that limit downside risk into the manager’s dynamic decision problem.

In practice, firms are often subject to risk management requirements imposed either internally by shareholders or externally by regulators. A popular risk measure in this context is Value-at-Risk (VaR), which limits the probability that the firm value falls below a prespecified floor. VaR is attractive in our setting because it provides a simple and economically interpretable way to model downside protection at the firm level. VaR has long been a widely used industry standard (\cite{duffie1997overview, jorion1997value, dowd1998beyond, anthony2000financial,basel2011}), and extensively employed in both financial and non-financial institutions (see, for example, the evidence in \cite{bodnar19981998}).
VaR is also widely studied in the literature and has been extensively applied in recent research in mathematical finance, including portfolio selection and optimization, as well as insurance risk management (see \cite{mi2023optimal, pirvu2007portfolio, li2013fuzzy}).

This paper studies a dynamic corporate risk management problem in which a risk-averse manager chooses costly effort and project volatility under fixed salary and option-based compensation. The manager's compensation makes the objective function non-concave in the terminal firm value, and a VaR constraint is imposed on it to capture downside risk control. This framework enables us to study how the VaR constraint and the compensation structure jointly affect managerial effort, risk-taking decisions, and the distribution of terminal firm value. We also compare optimal managerial behavior and firm value under the VaR constraint with those in the benchmark case without it.

Solving the manager’s optimization problem is technically challenging for three reasons. First, our problem is formulated in continuous time, and the VaR constraint creates time inconsistency, making the standard dynamic programming principle inapplicable. We therefore use the martingale approach (\cite{pliska1986stochastic, cox1989optimal, karatzas1998methods}), which reformulates the manager's dynamic strategy as a static optimization problem at the terminal time and derives intermediate strategies with the martingale property. Second, because the manager's compensation includes a call option, the objective function is non-concave. To handle this issue, we apply the concavification technique as in \cite{carpenter2000does, berkelaar2004optimal, bensoussan2015entrepreneurial} to first solve the optimization problem on the concave envelope and then recover the solution to the original problem. Third, the VaR constraint imposes a probabilistic restriction on the firm value, making it difficult to handle directly within the optimization problem. We address this issue by applying the quantile formulation method developed in \cite{schied2004neyman, carlier2006law, jin2008behavioral, xu2016note}, which enables us to express the terminal firm value explicitly in terms of the pricing kernel. The quantile formulation is commonly used when a risk-measure constraint is imposed on terminal wealth in portfolio management problems, for example, in \cite{cahuich2013quantile, wei2018risk, wei2021risk, mi2023optimal}. Combining these tools allows us to explicitly derive optimal effort, terminal firm value, and project choice.

Our paper is closely related to \cite{chen2024equivalence, nguyen2020nonconcave}. \cite{nguyen2020nonconcave} consider a non-concave investment problem for an insurer providing participating life insurance contracts under a VaR regulatory constraint. They derive explicit optimal terminal wealth and the corresponding investment stratigies. \cite{chen2024equivalence} consider a related insurer optimization problem with non-concave objective function under VaR, Expected Shortfall, Expected Discounted Shortfall, and Average Value-at-Risk constraints, emphasizing the equivalence of these risk measures. Our paper differs from both studies in three main respects. First, their non-concavity is generated by nonlinear insurance payoffs and limited liability, while ours arises from the manager's fixed salary and option-based compensation. Second, they focus on the insurer's terminal wealth allocation and investment strategies, whereas we additionally consider the manager's costly effort and choices of project risk, which directly affect the expected growth and volatility of the firm's value. Finally, the papers apply different methodologies. We combine concavification and quantile formulation with the martingale approach, whereas the two insurance studies obtain the terminal payoff through pointwise or piecewise Lagrangian method. Nevertheless, the terminal-value structures are mathematically related: they all involve nonlinear positive part payoffs subject to a budget constraint and a regulatory requirement, resulting in piecewise terminal-payoff structures. Therefore, our contribution mainly lies in embedding the VaR-constrained terminal-value problem in a dynamic corporate model and analyzing how the constraint affects managerial effort and project risk.    

Our main findings are as follows. First, our analysis shows that the VaR constraint significantly reshapes the distribution of terminal firm value by smoothing the value across states. Compared with the benchmark case, the VaR manager sacrifices some upside firm value in favorable states to satisfy the VaR requirement in intermediate states, while bankruptcy may still occur in sufficiently adverse states. Depending on the VaR floor and the tail probability, the optimal terminal firm value may exhibit binary, ternary, or quaternary patterns. Second, we find that when the VaR constraint is binding, the manager exerts more effort than the manager in the benchmark case. Third, the VaR constraint generally lowers bankruptcy probability; however, when the VaR floor is excessively high, the bankruptcy probability may increase. Finally, our sensitivity analysis shows that the optimal strategy depends significantly on both the manager’s productivity and the compensation design. Higher productivity uniformly improves the firm value across all non-bankruptcy states. Furthermore, a compensation package with more options encourages the manager to take more responsibility and smooth the terminal firm value across states. In contrast, a higher fixed salary makes managers less responsible and leads to a more dispersed firm value. Lastly, the effects of option incentives, risk aversion, and VaR constraints jointly create a non-monotonic optimal project choice.

This paper contributes to the literature in three aspects. First, it introduces a VaR-based risk management constraint into a dynamic managerial model with costly effort and option-based compensation, illustrating the effect of downside risk control on managerial effort, risk-taking actions, and the distribution of terminal firm value.
Second, despite the combination of a non-concave objective function and the VaR restriction, we obtain explicit solutions for the optimal firm value, effort level, and project volatility choices. Third, our numerical analysis provides economic insights into how the VaR constraint, managerial productivity, and compensation design shape managerial behavior and firm value.

The rest of the paper is organized as follows: Section \ref{sec:problem_formulation} presents the model setup and formulates the risk management problem. Section \ref{sec:optimal} derives the optimal strategies for the benchmark and VaR-constrained managers. Section \ref{sec:numerical} provides a sensitivity analysis and discusses the economic implications under CRRA utility. Section \ref{sec:conclusion} concludes. Proofs are collected in supplementary notes. 

\section{Problem formulation}\label{sec:problem_formulation}
In this section, we formulate our corporate risk management problem. We consider a firm whose value follows a controlled diffusion process. A risk-averse manager can exert costly effort and choose levels of volatility to influence the firm value dynamics. The manager derives utility from terminal compensation and experiences instantaneous disutility from exerting effort. In addition, the manager needs to control the risk of the firm, which is measured by the Value-at-Risk (VaR).

\subsection{Firm value model}
Let $T > 0$ be the given investment horizon and
$(\Omega,\mathcal{F},(\mathcal{F}_t)_{0 \le t \le T} ,\mathbb{P})$ be an atomless filtered probability space. $W(t)$ is a standard $\mathcal{F}_t$-adapted $1$-dimensional
Brownian motion with $W(0) = 0$. It is assumed that $\mathcal{F}_t = \sigma \{ W(s) : 0 \le s \le t \}$, augmented
by all the $\mathbb{P}$-null sets in $\mathcal{F}$.

Following \cite{cadenillas2004leverage,cadenillas2007optimal,bensoussan2015entrepreneurial}, we assume that the value of the firm is governed by the following stochastic differential equation (SDE):
\begin{equation}\label{eq:SDE_V}
\begin{aligned}
dV(t)&=r V(t)dt + \delta u(t)dt + \theta v(t) dt + v(t)dW(t), \\
V(0)&=V_0 >0.
\end{aligned}
\end{equation}
where $r>0, \delta > 0,\theta > 0$ are constant, $u$ and $v$ are adapted stochastic processes chosen by the manager. We assume
\begin{equation*}
\E\left[\int _0^T v^2(t)dt\right] < \infty \mbox{ and } \E\left[\int _0^T u^2(t) dt\right] < \infty ,
\end{equation*}
and $V(T) \ge 0$, which implies that the firm must settle all the liabilities by the terminal date. Following \cite{bensoussan2015entrepreneurial}, we do not impose the nonnegative firm value constraint for the entire period; the firm value $V(t)$ can be negative before $T$, as the manager can exert effort to restore the firm value. The model can be interpreted in the following way. $r$ is exogenous and stands for the baseline growth rate of firm value. $u$ is the level of effort the manager expends in running the company. $v$ represents the riskiness of the strategy chosen by the manager from a menu of projects with different levels of risk. The manager controls the firm’s volatility as well as the expected value through $v$. The parameter $\theta$ is a measure of the benefits associated with taking more risks. For example, $\theta$ will be high for those companies with stronger growth prospects, such as new firms. $\delta$ measures the impact of the manager’s effort on the value of the firm and can be interpreted as an
indicator of the manager's ability.

We define the following exponential martingale $Z(t)$
\begin{equation*}
Z (t):= \text{exp} \left\{ - (r + \frac{1}{2}\theta ^2) t - \theta W(t) \right\}.
\end{equation*}
$Z(t)$ is also known as the pricing kernel.
The following proposition, obtained by the martingale method, expresses the time-$t$ firm value as a function of the expectation of future value and the manager's effort.
\begin{prop}\label{prop:wealth}
The static budget constraint for the firm value is given by
\begin{equation}
V(t) = \frac{1}{Z (t) } \E \left[ Z (T) V(T) -  \delta \int _t^T Z (s) u (s)ds|\mathcal{F}_t\right], t \in [0,T].
\end{equation}
\end{prop}

\subsection{The manager's problem}
The manager exerts costly effort and chooses the volatility of the firm. In return, the manager receives a fixed salary and stock options as compensation. We assume the manager has no outside income. The objective of the manager is
\begin{equation*}
\max _{u,v} ~ \E \left[U(w + n(V(T) - K)^{+}) - \frac{1}{2}\int _0^T u^2(t)dt\right],
\end{equation*}
where $w > 0$ is the fixed salary, $n \in (0,1)$ is the number of shares of stock options or the percentage of positive profits, and $K$ is the strike price. This compensation structure is consistent with the managerial compensation literature. $U(\cdot)$ denotes the manager's utility function and satisfies the following standard assumptions. The manager's effort level, $u$, creates a disutility of $u^2/2$, and the project volatility $v$ implicitly affects the manager's utility.

\begin{Assumption}
$U: \mathbb{R}_{+} \to \mathbb{R}$ is strictly increasing and continuously   differentiable. Furthermore, $U
^{'}(\cdot)$ is strictly decreasing and satisfies the Inada condition, i.e., $U^{'}(0+) = +\infty $ and  $U^{'}(+\infty)=0$. $I(\cdot):= (U^{'})^{-1}(\cdot)$ satisfies the growth condition: there exists $K_1,K_2 >0$ such that $I(1/y)<K_1 y^{K_2}$ for all sufficiently large $y$.
\end{Assumption}

Inspired by the standard martingale approach (see \cite{pliska1986stochastic,cox1989optimal,karatzas1998methods}), we consider the following optimization problem as a benchmark

\begin{equation}\label{prob:B}
\begin{aligned}
\max _{V(T),u} ~ &\E \left[U(w + n(V(T) - K)^{+}) - \frac{1}{2}\int _0^T u^2(t) dt\right]\\
\text{subject to} ~ &\E \left[ Z (T) V(T) -  \delta \int _0^T Z (t) u(t)dt\right] \le V(0),\\
&V(T) \ge 0 . 
\end{aligned}
\end{equation}

To capture downside-risk control, we impose a VaR constraint on the terminal firm value. We follow \cite{basak2001value,wei2018risk} to embed risk management into \eqref{prob:B}, and the manager’s optimization problem becomes

\begin{equation}\label{prob:var}
\begin{aligned}
\max _{V(T),u} ~ &\E \left[U(w + n(V(T) - K)^{+}) - \frac{1}{2}\int _0^T u^2(t) dt\right]\\
\text{subject to} ~ &\E \left[ Z (T) V(T) -  \delta \int _0^T Z (t) u(t)dt\right] \le V(0), \\
& \P(V(T) \ge \underline{V}) \ge 1 - \alpha,\\
& V(T) \ge 0.
\end{aligned}
\end{equation}
where $\underline{V}>0$ is the floor value, and $\alpha$ is the tail probability, which are determined by shareholders or regulators. Note that the VaR constraint is equivalent to $\P(V(0) - V(T) \le V(0) - \underline{V}) \ge 1 - \alpha $, thus consistent with the literature, for example, \cite{duffie1997overview,jorion1997value}.
For convenience, we refer to managers without a risk constraint \eqref{prob:B} as benchmark managers, and managers with the VaR constraint \eqref{prob:var} as VaR managers.

\section{Optimal strategies}\label{sec:optimal}
\begin{figure}
    \centering
\includegraphics[width=0.8\linewidth]{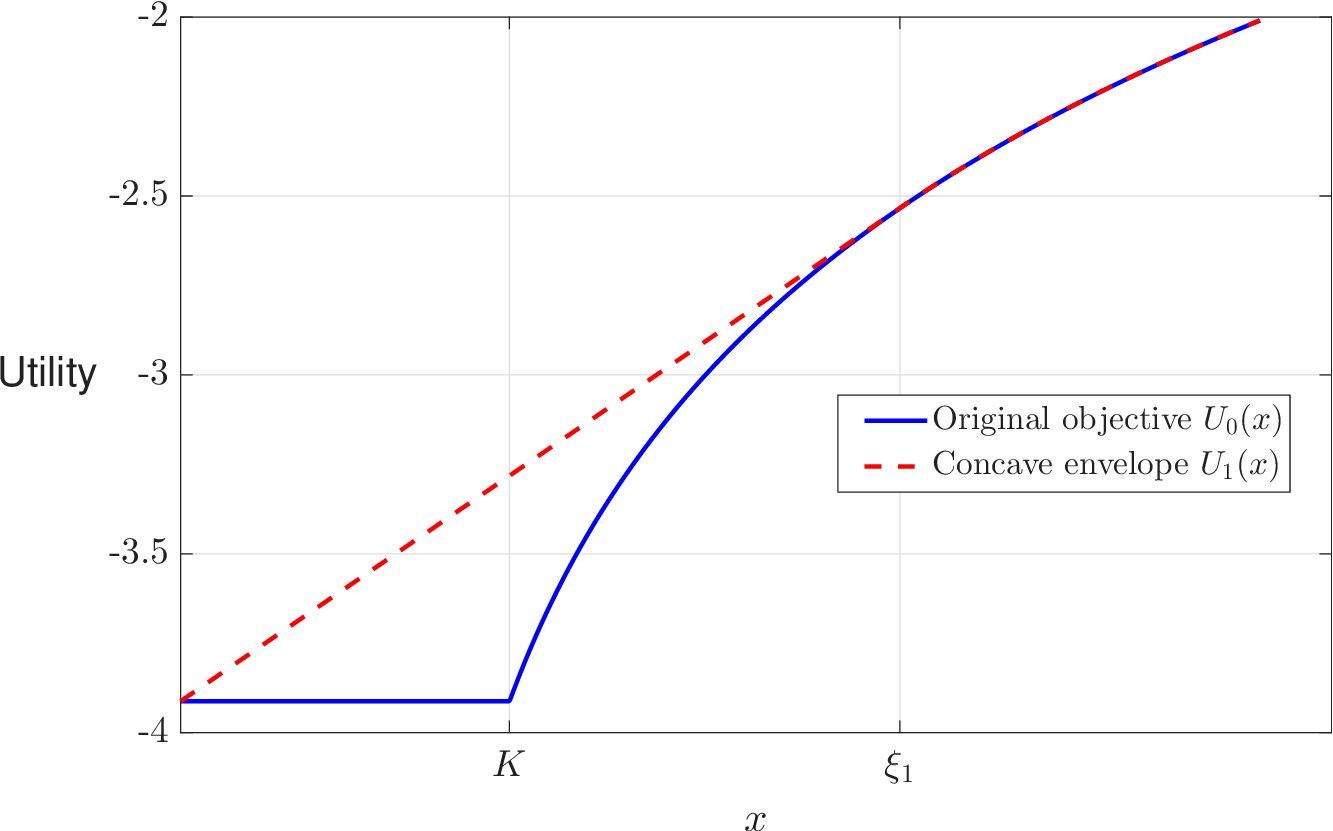}
    \caption{Concave Envelope}
    \label{fig:envelope}
\end{figure}

In this section, we derive solutions to problems \eqref{prob:B} and \eqref{prob:var}.
Because the objective function in both problems, $U_0(x) :=U(w + n(x - K)^{+})$, is  non-concave in $x$, we apply the concavification technique developed in \cite{carpenter2000does,berkelaar2004optimal,bensoussan2015entrepreneurial}. We introduce $U_1$, the concave envelope of $U_0$:
\begin{equation}\label{eq:U1}
	U_1 (x)=\left\{ 
	\begin{array}{ll}
U(w + n(x - K)^{+}) & ~   x \ge \xi_1,\\
		U(w) + n \beta_1 x & ~ 0 \le x < \xi_1 ,
  \end{array}
	\right.
	\end{equation}
where $\xi_1 >K $ is the unique solution to
\begin{equation}\label{eq:xi}
U(w) + n \beta_1\xi_1 = U(w + n(\xi_1 - K)^{+}),
\end{equation}
and $\beta_1 = U^{'}(w + n(\xi_1 - K)^{+})$.
Figure \ref{fig:envelope} plots the original non-concave objective function $U_0(x)$ and its concave envelope $U_1(x)$. The concave envelope is tangent to the original objective function at $\xi_1$.

We make the following integrability assumption.
\begin{Assumption}\label{assump:inte}
$\E \left[ \frac{1}{n}Z (T)  I\left(\frac{1}{n} \lambda Z (T)\right) -   \lambda \delta ^2 \int _0^T Z^2 (t)dt\right] < \infty$
for all $\lambda >0$.
\end{Assumption}

\subsection{Optimal strategy for benchmark managers}

Since the objective function in the benchmark problem \eqref{prob:B} is non-concave, we first consider the following auxiliary problem in which the objective is replaced by its concave envelope $U_1$. The resulting optimizer is also optimal for the original benchmark problem. The auxiliary problem is given by
\begin{equation}\label{prob:B_envelope}
\begin{aligned}
\max _{V(T),u} ~ &\E \left[U_1(V(T)) - \frac{1}{2}\int _0^T u^2(t) dt\right]\\
\text{subject to} ~ &\E \left[ Z (T) V(T) -  \delta \int _0^T Z (t) u(t)dt\right] \le V(0), \\
&V(T) \ge 0 . 
\end{aligned}
\end{equation}

The following proposition presents the optimal strategy of the benchmark manager, which is the solution to \eqref{prob:B}. The proof is analogous to that in \cite{bensoussan2015entrepreneurial} with minor modifications and is hence omitted.

\begin{prop}\label{prop:non-risk}
The optimal strategy of the benchmark manager is characterized as follows.
\begin{enumerate}
\item \textbf{The optimal effort.} The optimal effort is
\begin{equation*}
u_{0}(t) = \lambda _{0} \delta Z(t);
\end{equation*}

\item \textbf{Optimal terminal firm value.} The optimal time-T value of the firm is
\begin{equation}\label{eq:bench}
	V _{0}(T)=\left(K + \frac{1}{n} \left( I\left(\frac{1}{n} \lambda _0 Z (T)\right) - w\right)\right)\mathbf{1}_{\{Z(T)\le Z_0\}},
\end{equation}
	where $Z_0 = n\beta_1/\lambda _0$ and $\lambda _0 >0$ solves
	\begin{equation*}
	\E \left[ Z (T) V_0(T) -  \delta \int _0^T Z (t) u_0(t)dt\right] = V(0);
	\end{equation*}
 	
\item \textbf{Optimal project choice.} Let 
\begin{equation*}
 V_0(t) = \frac{1}{Z (t) } \E \left[ Z (T) V_0(T) -  \delta \int _t^T Z (s) u_0 (s)ds\Big|\mathcal{F}_t\right].
\end{equation*}
The optimal project choice is
\begin{equation*}
v_0(t) = - \theta \dfrac{\partial V_0(t)}{\partial Z(t)}Z(t).
\end{equation*}

\end{enumerate}
\end{prop}

\subsection{Optimal strategy for VaR managers}
For the VaR manager, we consider the following auxiliary problem with concavification.
\begin{equation}\label{prob:var_envelope}
\begin{aligned}
\max _{V(T),u} ~ &\E \left[U_1(V(T)) - \frac{1}{2}\int _0^T u^2(t) dt\right]\\
\text{subject to} ~ &\E \left[ Z (T) V(T) -  \delta \int _0^T Z (t) u(t)dt\right] \le V(0), \\
& \P(V(T) \ge \underline{V}) \ge 1 - \alpha,  \\
& V(T) \ge 0 . 
\end{aligned}
\end{equation}

Let $\mathbb{F}$ denote the set of cumulative distribution functions (CDFs) of all lower-bounded random variables taking values on $\mathbb{R}$, that is,
\begin{equation*}
\begin{aligned}
\mathbb{F} = \{F(\cdot) : \mathbb{R} \to [0, 1],\text{ non-decreasing, right continuous, }\\
F(a-) = 0 \text{ for some } a \in \mathbb{R} \text{ and } F(+\infty ) = 1\}.
\end{aligned}
\end{equation*}
For any $F( \cdot ) \in  \mathbb{F}$, denote by $F^{-1}(\cdot)$ its inverse function, that is,
\begin{equation*}
F^{-1}(t) = \inf \{x \in \mathbb{R} : F(x) \geq t \} = \sup \{x\in \mathbb{R} : F(x) < t \}, ~t \in [0, 1].
\end{equation*}

Denote by $F_{Z}(\cdot)$ the cumulative distribution function of $Z(T)$, and $F^{-1}_{Z}(\cdot)$ the quantile function of $Z(T)$. We first solve problem \eqref{prob:var_envelope} and then identify the region in which the resulting solution takes values where the concave envelope differs from the original objective. After that, we reanalyze that part under the original objective. This yields the solution to problem \eqref{prob:var}.
The following proposition presents the optimal strategy for VaR managers. Specifically, the optimal terminal firm value is separated into 9 cases depending on the VaR floor $\underline{V}$ and the tail probability $\alpha$.

\begin{prop}\label{prop:var}
The optimal strategy of the VaR manager is characterized as follows.
\begin{enumerate}
\item \textbf{The optimal effort.}
\begin{equation*}
u_{VaR}(t) = \lambda _{\text{VaR}} \delta Z(t).
\end{equation*}

\item \textbf{Optimal terminal firm value.} The optimal terminal firm value $V_{\text{VaR}}(T)$ is given by the following 9 cases.
\begin{enumerate}
    \item $\underline{V}<K,\alpha<\alpha_1(\lambda_{\text{VaR}},\beta_2)$,
    \begin{equation}
	V_{\text{VaR}}(T):=\left\{ 
	\begin{array}{ll}
K + \frac{1}{n} \left( I\left(\frac{1}{n} \lambda_{\text{VaR}} Z(T)\right) - w\right) & ~ 0<Z(T)\le Z_{\text{VaR}}(\beta_2) ,\\
\underline{V} & ~  Z_{\text{VaR}}(\beta_2)<  Z(T) \le Z_{\text{VaR},1},\\
		0 & ~  Z(T) > Z_{\text{VaR},1};
  \end{array}
	\right.
\end{equation}
\item $\underline{V}<K,\alpha_1(\lambda_{\text{VaR}},\beta_2)\le \alpha < \alpha_1(\lambda_{\text{VaR}},\beta_1),$
\begin{equation}
	V_{\text{VaR}}(T):=\left\{ 
	\begin{array}{ll}
K + \frac{1}{n} \left( I\left(\frac{1}{n} \lambda_{\text{VaR}} Z(T)\right) - w\right)  & ~   0< Z(T) \le Z_{\text{VaR},1},\\
		0 & ~  Z(T) > Z_{\text{VaR},1};
  \end{array}
	\right.
\end{equation}

 \item $\underline{V}<K,\alpha\ge \alpha_1(\lambda_{\text{VaR}},\beta_1)$, $V_{\text{VaR}}(T)=V_0(T)$ and $\lambda_{\text{VaR}}=\lambda_0$;

\item $K\le\underline{V}<\xi_1,\alpha<\alpha_2(\lambda_{\text{VaR}})$,
\begin{equation}
	V_{\text{VaR}}(T):=\left\{ 
	\begin{array}{ll}
K + \frac{1}{n} \left( I\left(\frac{1}{n} \lambda_{\text{VaR}} Z(T)\right) - w\right)  & ~  0<Z(T) \le Z_{\text{VaR},2},\\
\underline{V} & ~ Z_{\text{VaR},2} < Z(T) \le Z_{\text{VaR},1},\\
		0 & ~ Z(T)> Z_{\text{VaR},1};
  \end{array}
	\right.
\end{equation}

\item  $K\le\underline{V}<\xi_1,\alpha_2(\lambda_{\text{VaR}})\le \alpha<\alpha_1(\lambda_{\text{VaR}},\beta_1),$
\begin{equation}
	V_{\text{VaR}}(T):=\left\{ 
	\begin{array}{ll}
K + \frac{1}{n} \left( I\left(\frac{1}{n} \lambda_{\text{VaR}} Z(T)\right) - w\right)  & ~  0< Z(T) \le Z_{\text{VaR},1} ,\\
		0 & ~  Z(T) > Z_{\text{VaR},1};
  \end{array}
	\right.
\end{equation}
\item $K\le\underline{V}<\xi_1, \alpha\ge\alpha_1(\lambda_{\text{VaR}},\beta_1),$ $V_{\text{VaR}}(T)=V_0(T)$ and $\lambda_{\text{VaR}}=\lambda_0$;

\item $\underline{V} \ge\xi_1, \alpha < \alpha_1(\lambda_{\text{VaR}},\beta_1)$, 
\begin{equation*}
	V _{\text{VaR}}(T)=\left\{ 
	\begin{array}{ll}
K + \frac{1}{n} \left( I\left(\frac{1}{n} \lambda_{\text{VaR}} Z(T)\right) - w\right)  & ~ 0< Z(T) \le Z_{\text{VaR},2} ,\\
\underline{V} & ~ Z_{\text{VaR},2} <  Z(T) \le Z_{\text{VaR},1},\\
		0 & ~ Z(T) > Z_{\text{VaR},1}  ;
  \end{array}
	\right.
\end{equation*}

\item $\underline{V} \ge \xi_1,  \alpha_1(\lambda_{\text{VaR}},\beta_1)\leq \alpha < \alpha_2(\lambda_{\text{VaR}})$, \begin{equation*}
	V _{\text{VaR}}(T)=\left\{ 
	\begin{array}{ll}
	K + \frac{1}{n} \left( I\left(\frac{1}{n} \lambda_{\text{VaR}} Z(T)\right) - w\right)  & ~ 0<  Z(T) \le Z_{\text{VaR},2} ,\\
	\underline{V} & ~ Z_{\text{VaR},2} <  Z(T) \le Z_{\text{VaR},1},\\
K + \frac{1}{n} \left( I\left(\frac{1}{n} \lambda_{\text{VaR}} Z(T)\right) - w\right)  & ~ Z_{\text{VaR},1} <  Z(T) \le Z_{\text{VaR}}(\beta_1), \\
0 & ~ Z(T) > Z_{\text{VaR}}(\beta_1) ;
  \end{array}
	\right.
\end{equation*}

\item $\underline{V} \ge \xi_1,  \alpha \ge \alpha_2(\lambda_{\text{VaR}}),$ $V_{\text{VaR}}(T)=V_0(T)$ and $\lambda_{\text{VaR}}=\lambda_0$;
\end{enumerate}

where
\begin{equation*}
\begin{aligned}
\alpha_1(\lambda,\beta)&:=1-F_Z\left(\dfrac{\beta}{\lambda}n\right),\\
\alpha_2(\lambda)&:= 1  - F_Z \left(\frac{n}{\lambda}U^{'}(w + n(\underline{V}-K)^+)\right),\\
Z_{\text{VaR},1} &:= F_Z ^{-1}(1-\alpha),\\
Z_{\text{VaR},2} &:= \frac{n}{\lambda _{\text{VaR}}} U^{'}(w + n( \underline{V} - K)^{+}) ,\\
Z _{\text{VaR}}(\beta) &:= \frac{n\beta}{\lambda _{\text{VaR}}} ,\\
\beta_2&=U^\prime(w+n(\xi_2-K)),
\end{aligned}
\end{equation*}
$\xi_2$ is the unique solution to
\[
U(w)+n\beta_2(\xi_2-\underline{V})=U(w+n(\xi_2-K)^+), \quad \underline{V}<K,
\]

and $\lambda _{\text{VaR}} $  solves
\begin{equation*}
\E \left[ Z (T) V_{\text{VaR}}(T) -  \delta \int _0^T Z (t) u_{\text{VaR}}(t)dt\right] = V(0).
\end{equation*}
\item \textbf{Optimal project choice.} Let 
\begin{equation*}
 V_{\text{VaR}}(t) = \frac{1}{Z (t) } \E \left[ Z (T) V_{\text{VaR}}(T) -  \delta \int _t^T Z (s) u_{\text{VaR}} (s)ds\bigg|\mathcal{F}_t\right].
\end{equation*}
The optimal project choice is
\begin{equation*}
v_{\text{VaR}}(t) = - \theta  \dfrac{\partial V_{\text{VaR}}(t)}{\partial Z(t)} Z(t).
\end{equation*}

    \item \textbf{Comparison with the benchmark.} If the VaR constraint is binding, then
    $\lambda _{\text{VaR}} > \lambda _0$, and thus $u_{\text{VaR}}(t) > u_{0}(t)$.
    If the VaR constraint is inactive, then
    $\lambda_{\text{VaR}}=\lambda_0$ and $u_{\text{VaR}}(t)=u_0(t)$.
    
\end{enumerate}
\end{prop}

The last statement in Proposition \ref{prop:var} shows that the VaR manager will exert more effort than the benchmark manager. The following Corollary provides a unified representation of the nine cases of the optimal terminal value, which can be used to solve the Lagrange multiplier efficiently in numerical calculations. We define
\begin{equation}\label{eq:Ztilde}
\widetilde{Z}_{\text{VaR}}:=Z_{\text{VaR}}(\beta_2)\times\mathbf{1}_{\{\underline{V}<K\}} + Z_{\text{VaR},2} \times \mathbf{1}_{\{K\le\underline{V}<\xi_1\}}+Z_{\text{VaR}}(\beta_1)\mathbf{1}_{\{\underline{V}\ge \xi_1\}}.
\end{equation}

\begin{coro}\label{coro:V_combined}
    The optimal terminal value $V_{\text{VaR}}(T)$ in Proposition \ref{prop:var} can be written as
    \begin{align*}
      V_{\text{VaR}}(T)=&  \max\left\{K + \frac{1}{n} \left( I\left(\frac{1}{n} \lambda_{\text{VaR}} Z(T)\right) - w\right), \underline{V}\right\}\times\mathbf{1}_{\{Z(T)<\min(Z_{\text{VaR},1},\widetilde{Z}_{\text{VaR}})\}} \\
      &+\left(K + \frac{1}{n} \left( I\left(\frac{1}{n} \lambda_{\text{VaR}} Z(T)\right) - w\right)\right)\times\mathbf{1}_{\{Z_{\text{VaR},1}\le Z(T)<Z_{\text{VaR}}(\beta_1)\}}\\
      &+\underline{V}\times\mathbf{1}_{\{\widetilde{Z}_{\text{VaR}}\le Z(T)<Z_{\text{VaR},1}\}}.
    \end{align*}
\end{coro}

\begin{coro}\label{coro:bankruptcy_prob}
When the VaR constraint is binding, the bankruptcy probability satisfies the following properties.
\begin{enumerate}
    \item When $\underline{V}<\xi_1$, $\mathbb{P}(V_{\text{VaR}}(T)=0)<\mathbb{P}(V_0(T)=0)$.
    \item When $\underline{V}\ge \xi_1$, $\mathbb{P}(V_{\text{VaR}}(T)=0)<\mathbb{P}(V_0(T)=0)$ if $\alpha<1-F_Z\left(\dfrac{\beta_1}{\lambda_0}n\right)$, and $\mathbb{P}(V_{\text{VaR}}(T)=0)\ge\mathbb{P}(V_0(T)=0)$ if $\alpha\ge1-F_Z\left(\dfrac{\beta_1}{\lambda_0}n\right)$.
\end{enumerate}
\end{coro}

The two corollaries can be verified by direct calculation; hence, the proof is omitted.

\section{Numerical analysis} \label{sec:numerical}
In this section, we compare the optimal strategies of the benchmark manager and the VaR managers. This section illustrates the economic implications of the optimal strategies derived in Section \ref{sec:optimal}. We provide examples of the optimal terminal firm values under different combinations of $\underline{V}$ and $\alpha$, assess the impact of several model parameters on firm value, and analyze the project volatility choices. We specify our analysis under the CRRA utility
\[
U(X) = \dfrac{X^{1-\gamma}}{1-\gamma} ,
\]
where
$0<\gamma\neq1$ is the relative risk aversion parameter.

Given the compensation parameters $(w, n, K)$ and the risk aversion parameter $\gamma$, we first solve for the tangency point $\xi_1$ of the global concave envelope $U_1$ and calculate the slope $\beta_1$ accordingly. When $\underline{V}<K$, the concavification technique used in Proposition \ref{prop:var} also requires the auxiliary tangency point $\xi_2$ and the corresponding slope $\beta_2$. Next, we solve the budget constraints for the benchmark and the VaR managers, obtaining $\lambda_0$ and $\lambda_{\text{VaR}}$. All case-specific thresholds, including $Z_{\text{VaR},1},Z_{\text{VaR},2}, Z_{\text{VaR}}(\beta_1), \text{ and }Z_{\text{VaR}}(\beta_2)$, follow immediately. Then the optimal strategies can be solved and visualized as a function of the pricing kernel $Z(T)$. The baseline parameters are
\[
T=1, r=0.05, \theta=0.4, V(0)=1,\delta=0.1, K=1,n=0.05,w=0.02, \gamma=1.5.
\]
With the baseline parameters, we have $\xi_1=1.8194$ by solving Equation \eqref{eq:xi}.

\subsection{Optimal terminal firm value}
\begin{figure}
    \centering
    \includegraphics[width=0.8\linewidth]{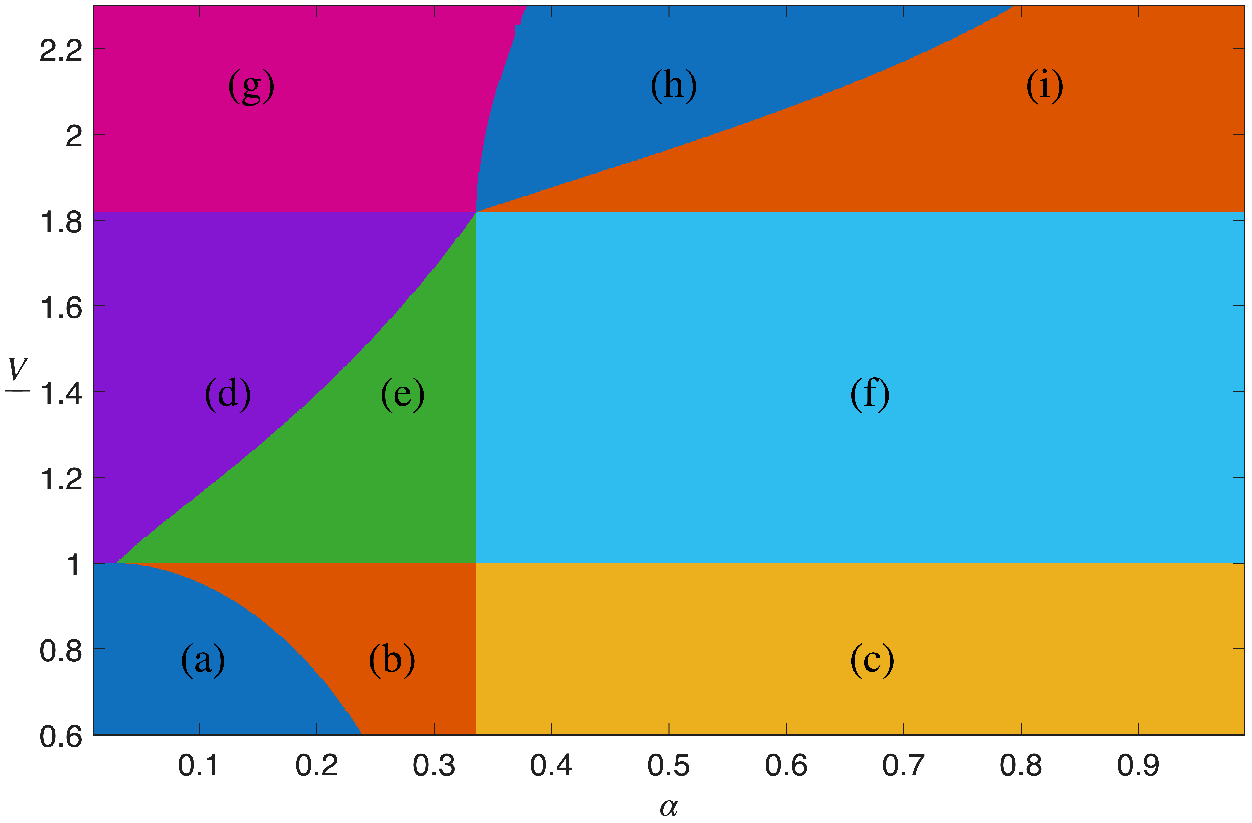}
    \caption{9 Regimes based on $(\underline{V},\alpha)$}
    \label{fig:regimemap}
\end{figure}

\begin{figure}[!htbp]
    \centering
    
    \begin{subfigure}[b]{0.48\textwidth}
        \centering
    \includegraphics[width=\textwidth]{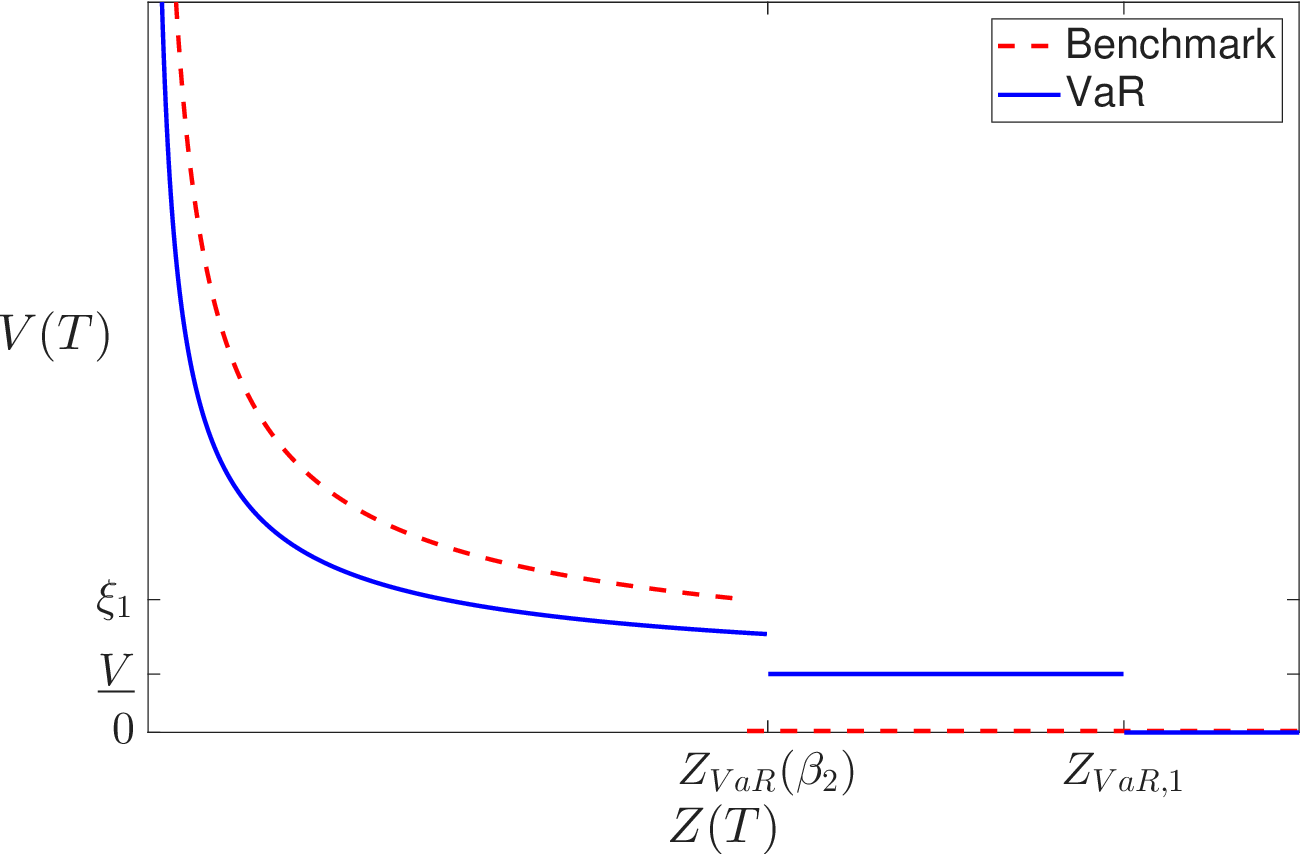}
        \caption{Case (a): $\underline{V} < K$, tight VaR}
        \label{fig:case_a}
    \end{subfigure}
    \hfill
    \begin{subfigure}[b]{0.48\textwidth}
        \centering
        \includegraphics[width=\textwidth]{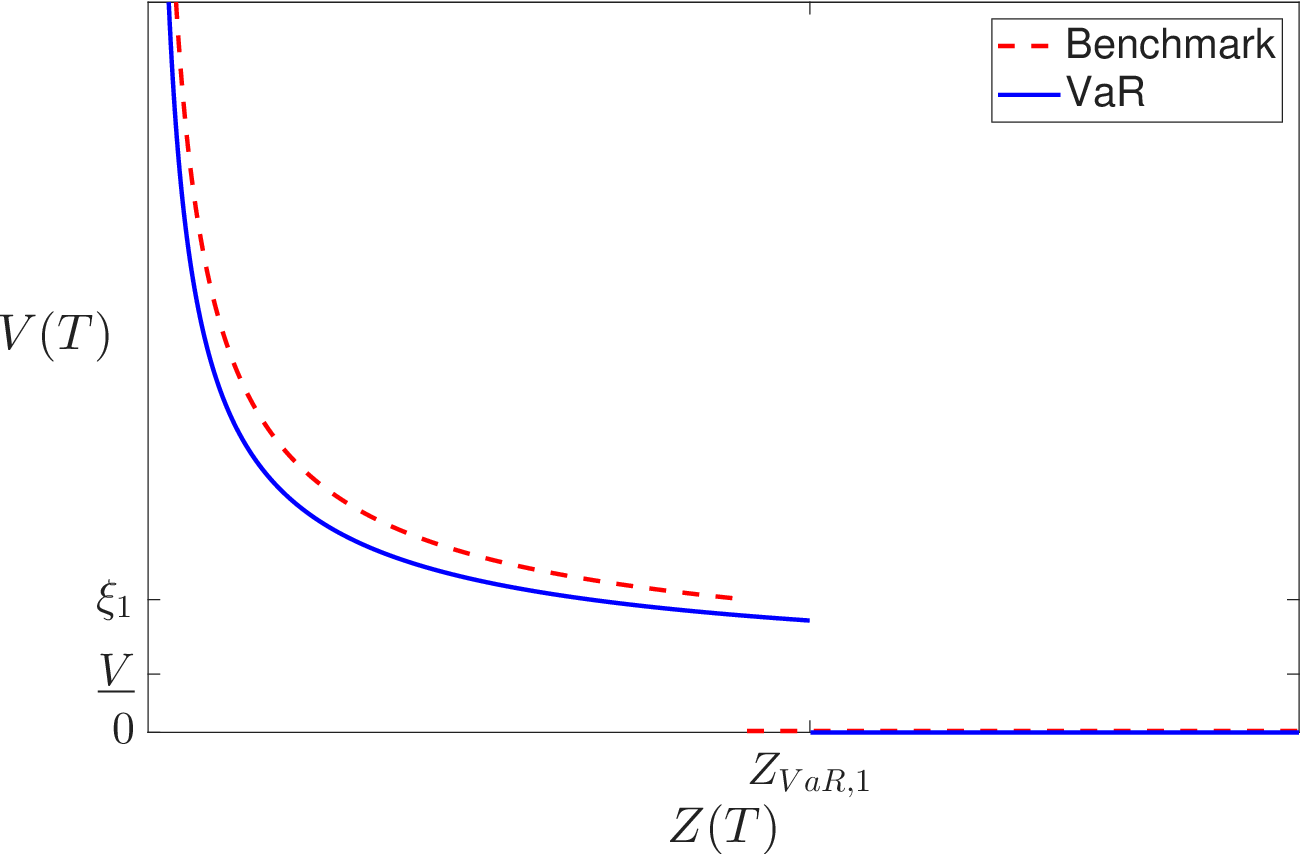}
        \caption{Case (b): $\underline{V} < K$, moderate VaR}
        \label{fig:case_b}
    \end{subfigure}

    \vspace{0.5em}

    \begin{subfigure}[b]{0.48\textwidth}
        \centering
        \includegraphics[width=\textwidth]{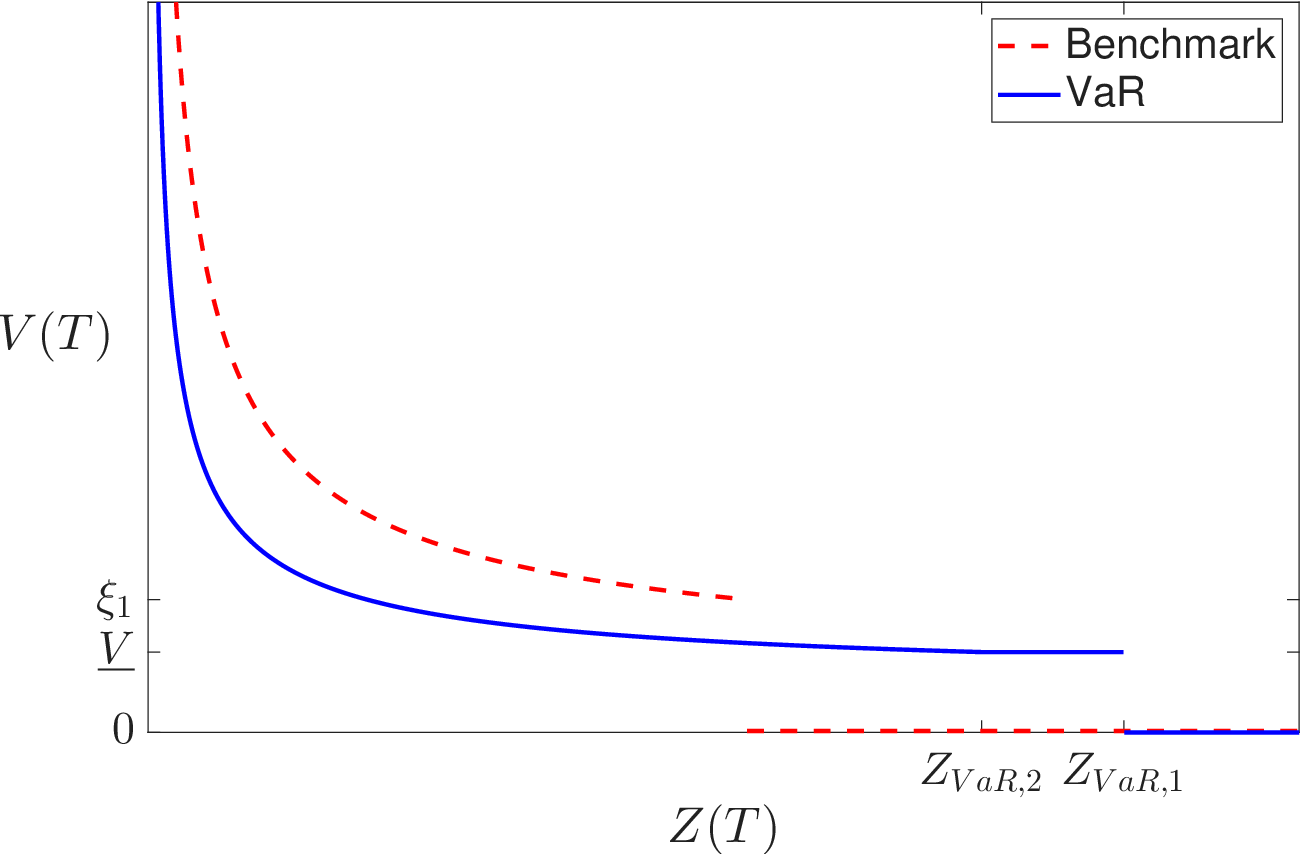}
        \caption{Case (d): $K \le \underline{V} < \xi_1$, tight VaR}
        \label{fig:case_d}
    \end{subfigure}
    \hfill
    \begin{subfigure}[b]{0.48\textwidth}
        \centering
        \includegraphics[width=\textwidth]{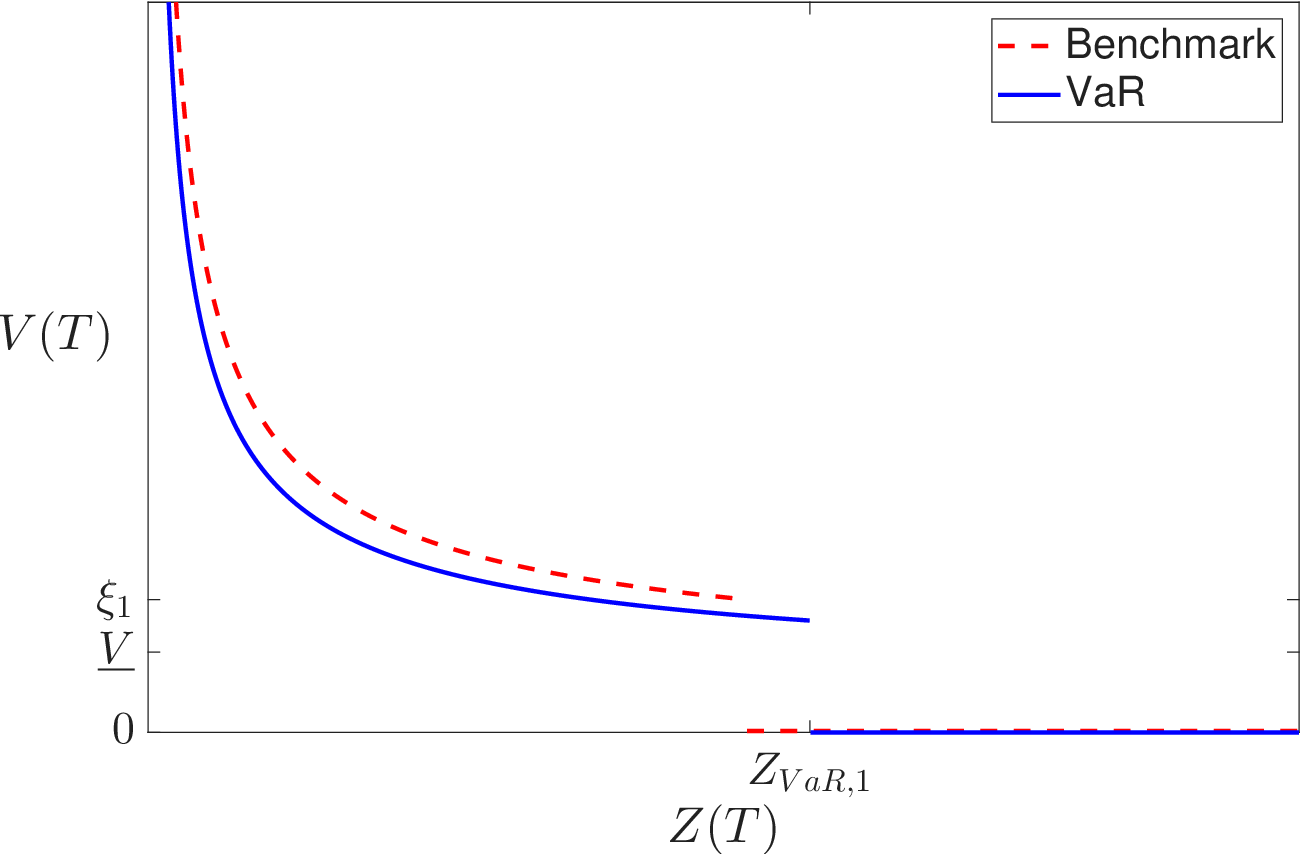}
        \caption{Case (e): $K \le \underline{V} < \xi_1$, moderate VaR}
        \label{fig:case_e}
    \end{subfigure}

    \vspace{0.5em}

    \begin{subfigure}[b]{0.48\textwidth}
        \centering
        \includegraphics[width=\textwidth]{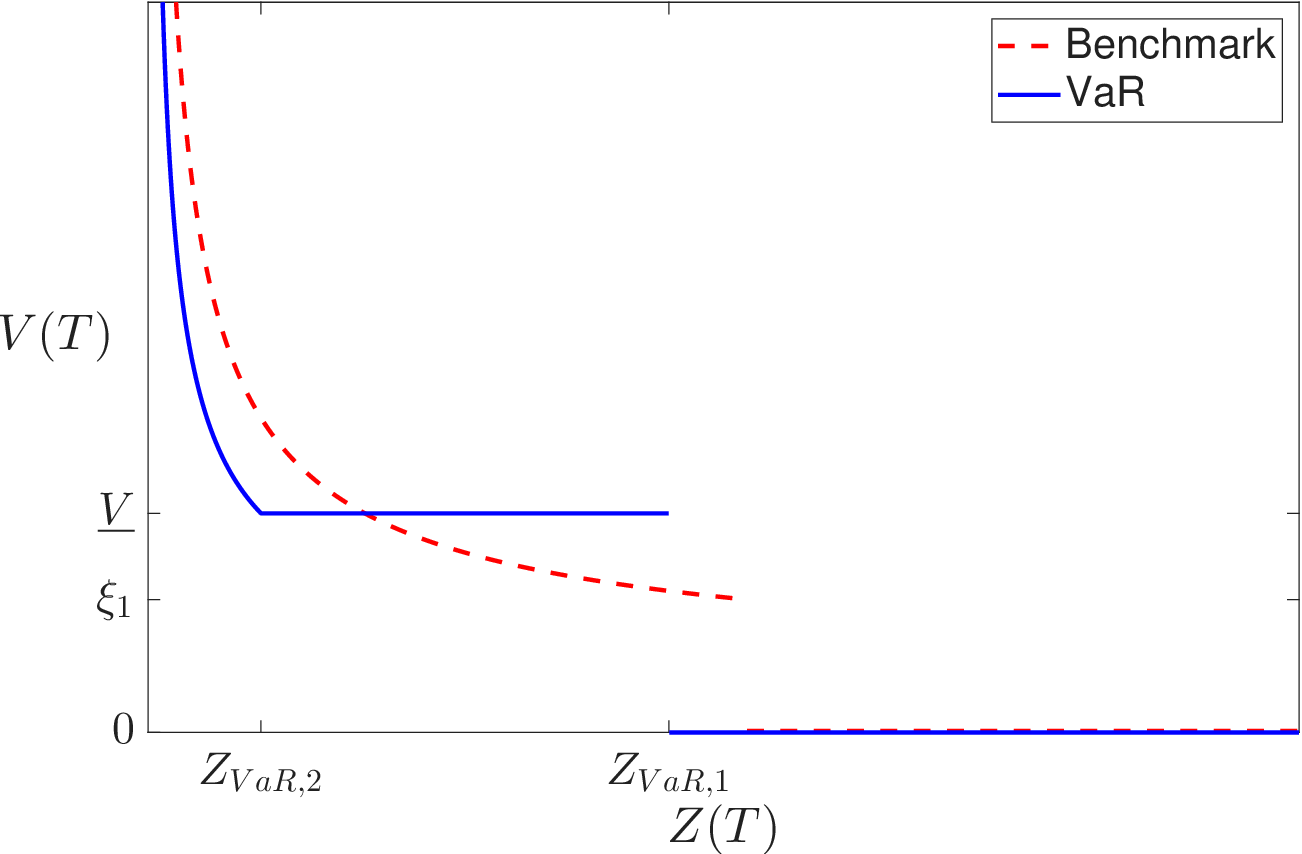}
        \caption{Case (g): $\underline{V} \ge\xi_1$, tight VaR}
        \label{fig:case_g}
    \end{subfigure}
    \hfill
    \begin{subfigure}[b]{0.48\textwidth}
        \centering
        \includegraphics[width=\textwidth]{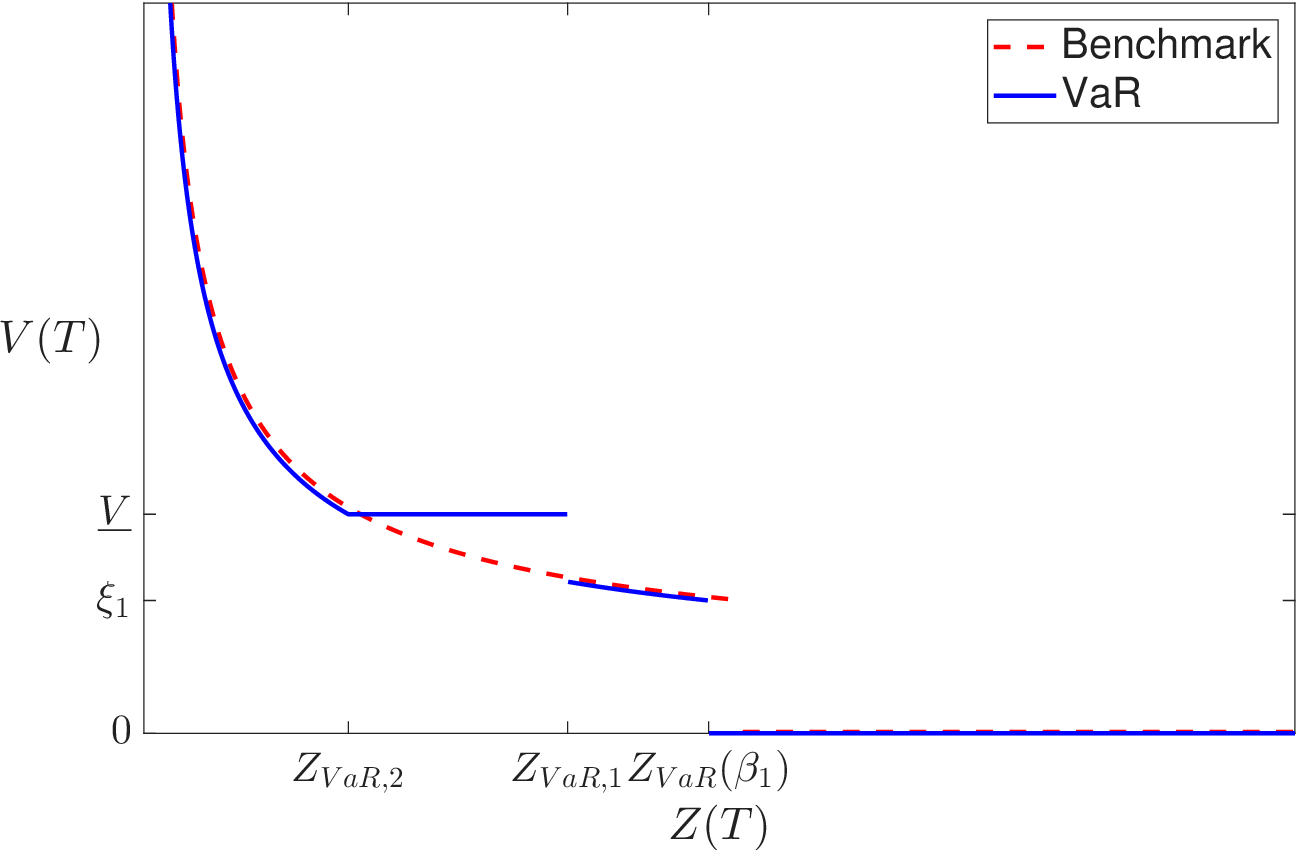}
        \caption{Case (h): $\underline{V} \ge \xi_1$, moderate VaR}
        \label{fig:case_h}
    \end{subfigure}

    \caption{Optimal terminal firm value in the six VaR-active regimes. Each panel plots the benchmark payoff \(V_0(T)\) and the VaR-constrained payoff \(V_{\text{VaR}}(T)\) as functions of the pricing kernel \(Z(T)\), for a representative parameterization of one of the six cases in Proposition \ref{prop:var} in which the VaR constraint is active.}
    \label{fig:6cases}
\end{figure}

Proposition \ref{prop:var} shows that the optimal terminal firm value can be divided into 9 cases depending on the choices of $(\underline{V},\alpha)$. Figure \ref{fig:regimemap} illustrates the partition of these 9 regimes in the $(\underline{V},\alpha)$ plane. We also plot the firm value of the six regimes in Proposition \ref{prop:var} in which the VaR constraint is active. Different combinations of $(\underline{V},\alpha)$ are chosen to generate the six cases. 
 Figure \ref{fig:6cases} collects one representative example from each of the six cases and plots the optimal terminal firm value of the benchmark manager, $V_0(T)$, and that of the VaR manager, $V_{\text{VaR}}(T)$, as functions of the pricing kernel $Z(T)$. The three remaining cases are omitted because the VaR constraint is inactive and the optimal payoff is equivalent to the benchmark solution.

The six panels in Figure \ref{fig:regimemap} show that the VaR constraint changes the manager's strategy through a common pattern of cross-state firm value reallocation. Relative to the benchmark manager, the VaR manager gives up part of the firm value in good states ($Z(T)$ is small) in order to satisfy the VaR requirement over some intermediate states ($Z(T)$ takes intermediate values). In the worst states, the manager may choose to make the firm bankrupt. As a result, the benchmark manager’s binary payoff pattern is replaced by a more complex piecewise structure with the VaR requirement. Depending on the value of the floor $\underline{V}$ and the tail probability $\alpha$, the VaR-constrained payoff can be binary, ternary, or quaternary. Intuitively, a low $\underline{V}$ and a high $\alpha$ indicate a less strict VaR constraint, and the VaR manager's strategy is close to or the same as the benchmark strategy. In contrast, a high $\underline{V}$ or a low $\alpha$ represents a stricter VaR constraint, and the constraint significantly affects the firm value.

The first two panels correspond to the case $\underline{V}<K$. That is, the VaR floor lies below the option strike. When the constraint is sufficiently tight (Case (a)), the optimal payoff exhibits a protected region at the floor $\underline{V}$ over intermediate states, while it still drops to zero in bad states. When the constraint is weaker (Case (b)), this protected region disappears, and the VaR payoff becomes a downward-shifted version of the benchmark payoff.

The next two panels correspond to $K\le \underline{V}<\xi_1$. The floor is between the strike and the tangency point. We still observe a similar pattern: a strict VaR constraint (Case (d)) generates a flat segment at $\underline{V}$, whereas a loose constraint (Case (e)) produces a binary payoff without the flat part at $\underline{V}$. In addition, in Case (d), the decreasing branch is connected to the flat segment ($Z_{\text{VaR},2}<Z(T)<Z_{\text{VaR},1}$), while in Case (a), there is a gap between the curved branch and the flat part.

The final two panels correspond to $\underline{V}>K$. This is the most restrictive region since the VaR floor lies above the tangency point. The flat region at $\underline{V}$ appears in both Cases (g) and (h). Additionally, the optimal firm value can become quaternary for intermediate values of $\alpha$ (Case (h)): the manager preserves $\underline{V}$ over some states, reverts to the curve-shaped strategy in worse states when the VaR constraint is no longer effective, and still defaults in sufficiently bad states. Also, since the terminal firm value of the benchmark manager can only take values above $\xi_1$ or 0, the flat segment at $\underline{V}>\xi_1$ of the VaR manager may exceed the curved part of the benchmark manager in some states.

\begin{figure}
    \centering
    \includegraphics[width=0.7\linewidth]{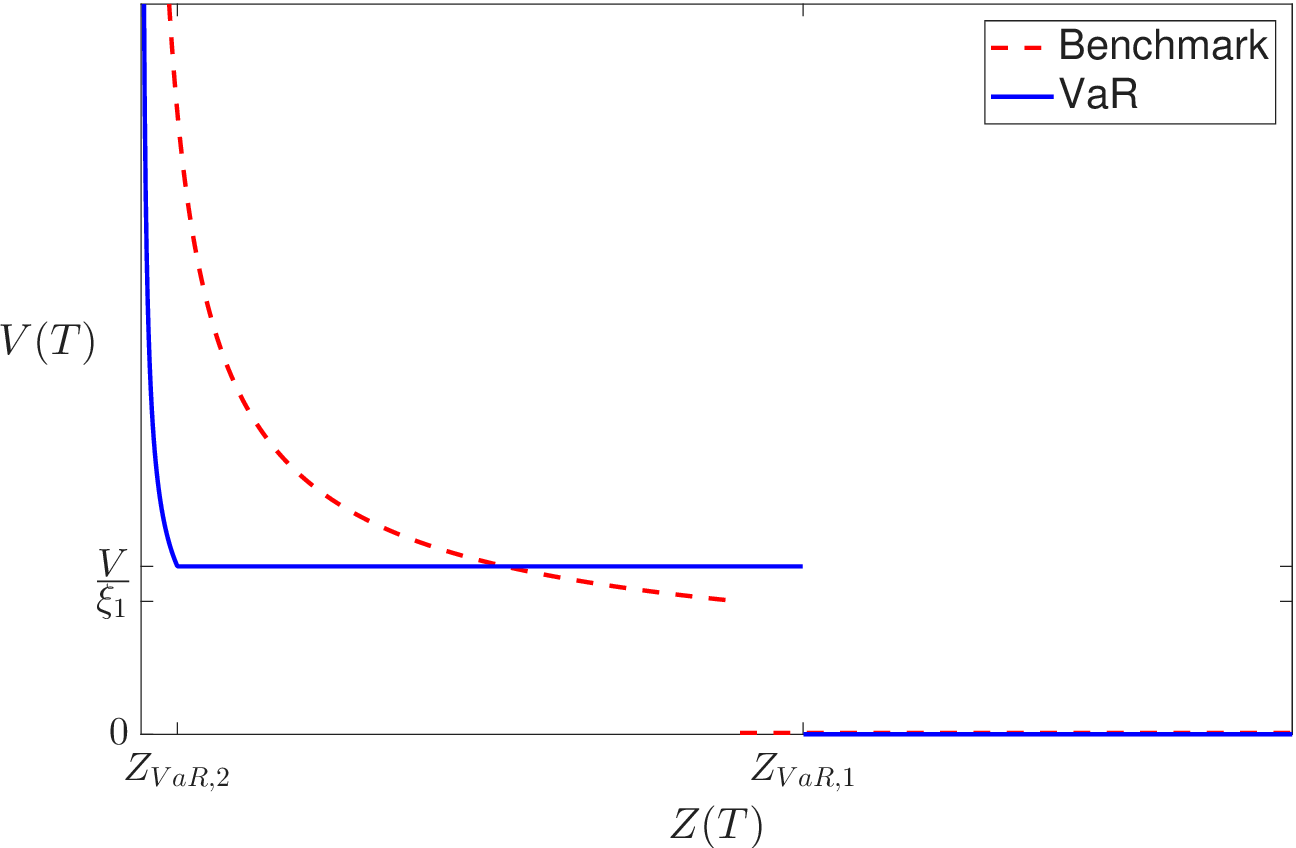}
    \caption{Case (g): $\underline{V}\ge\xi_1,\mathbb{P}(V_{\text{VaR}}(T)=0)<\mathbb{P}(V_0(T)=0)$}
    \label{fig:extra}
\end{figure}

Both Cases (g) and (h) in Figure \ref{fig:6cases} show that the default probability with the VaR constraint is higher than that of the benchmark case. To complement Figure \ref{fig:6cases}, we present another example for Case (g) in Figure \ref{fig:extra}, in which the VaR constraint lowers the probability of bankruptcy. In this example, the VaR manager chooses a smaller default region than the benchmark manager, so that $\mathbb{P}(V_{\text{VaR}}(T)=0)<\mathbb{P}(V_0(T)=0)$.
This pattern is consistent with the result in Corollary \ref{coro:bankruptcy_prob}. Our analysis indicates that the bankruptcy probability of the VaR manager is always lowered for $\underline{V}<\xi_1$ if the VaR constraint is active. However, for $\underline{V}\ge\xi_1$, the effect on bankruptcy is dependent on the choice of $\alpha$. Small $\alpha$ generates higher payoffs in bad states, reducing the default probability, while large $\alpha$ increases payoffs in intermediate states, leaving more bad states with bankruptcy.

The above analysis reveals that the VaR constraint can generate distinctive return patterns. In general, VaR reduces terminal firm value in extremely good states in order to support outcomes in less favorable states, but it does not eliminate bankruptcy in sufficiently bad states. When the constraint is tight enough, it creates a protected region at the floor $\underline{V}$ over intermediate states. VaR serves as a tool to not only manage the firm's risk but also generate new return patterns that are rare in the conventional case, better aligning with shareholders' risk preferences.


\subsection{Sensitivity to model parameters}
In this section, we conduct a sensitivity analysis of three parameters: the effort-productivity parameter $\delta$, the shares of stock option compensation $n$, and the fixed salary $w$. To avoid excessive repetition, we focus on the representative regimes in cases (d) and (e), where $K\le\underline{V}<\xi_1$ and the choice of $\alpha$ makes the VaR constraint binding. There are two cutoff points for the state variable $Z(T)$: $Z_{\text{VaR},1}$ and $Z_{\text{VaR},2}$. 
$Z_{\text{VaR},1}$ is the bankruptcy threshold, and the manager chooses to default whenever $Z(T)\ge Z_{\text{VaR},1}:=F_Z^{-1}(1-\alpha)$. Since $Z_{\text{VaR},1}$ only depends on $\alpha$, the three parameters do not affect the bankruptcy threshold. $Z_{\text{VaR},2}$ is the cutoff point for the VaR constraint to be effective. If $Z_{\text{VaR},1}>Z_{\text{VaR},2}$, the firm value takes the curved branch on states $Z(T)<Z_{\text{VaR},2}$ and takes the constant $\underline{V}$ on states $Z_{\text{VaR},2}\le Z(T)<Z_{\text{VaR},1}$. If $Z_{\text{VaR},1}\le Z_{\text{VaR},2}$, the constant part disappears and the firm value takes the curved branch on all states $Z(T)<Z_{\text{VaR},1}$.

The sensitivity analysis results reveal a significant distinction among the three parameters.  The parameter $\delta$ changes the productivity with which managerial effort is transformed into firm value, and it shifts the terminal firm value upwards uniformly across all the non-bankruptcy states. In contrast, $n$ and $w$ are compensation parameters. They alter how utility is generated from the firm's performance, and thereby reshape the payoff across states under the same VaR and budget constraints, typically increasing firm value in some states while decreasing it in others.

\begin{figure}[htbp]

    \centering
    
    \begin{subfigure}[b]{0.5\textwidth}
    \hspace*{-1.2cm}
    \includegraphics[width=3.7in,height=2.5in]{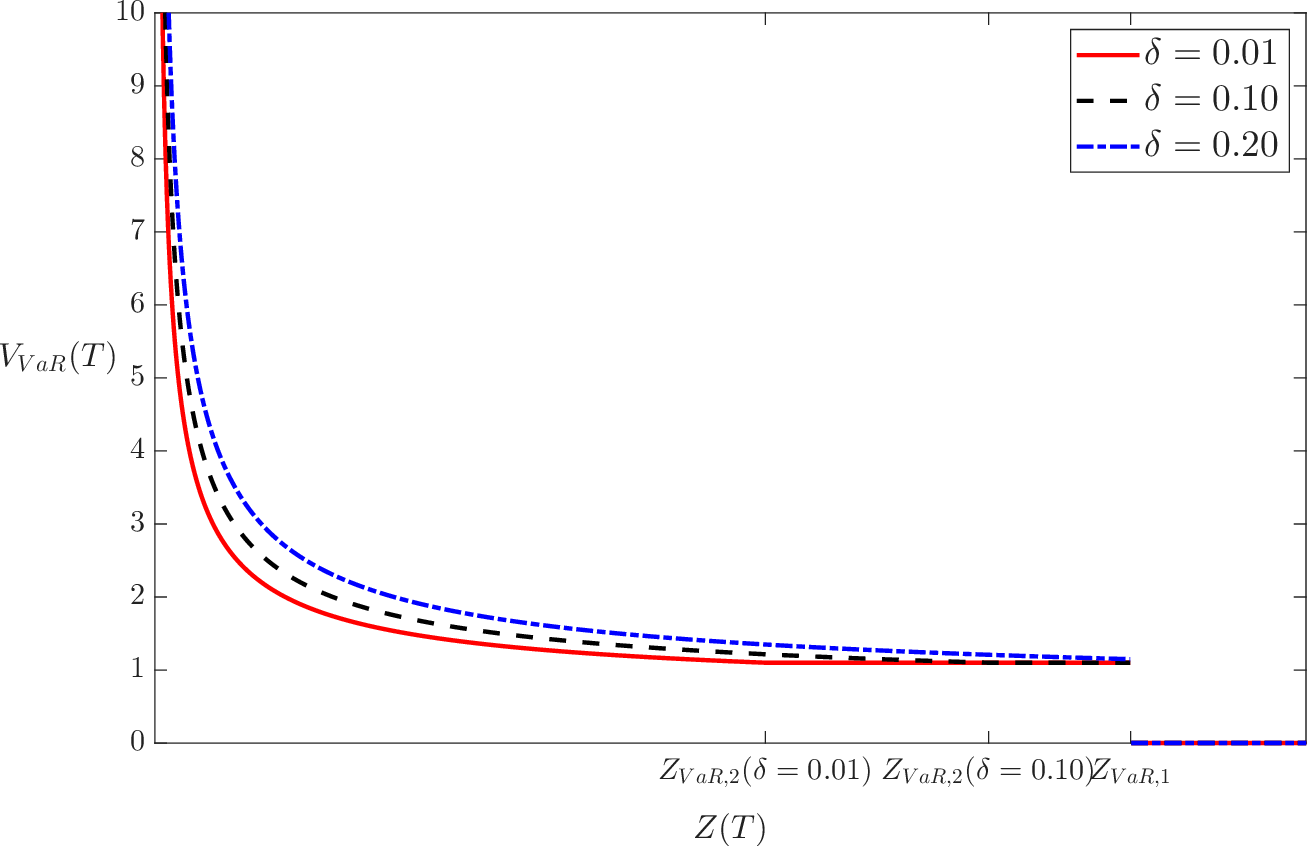}
        \caption{Sensitivity with $\delta$}
        \label{fig:sensi_delta}
    \end{subfigure}\\

    \begin{subfigure}[b]{0.5\textwidth}
    \hspace*{-1.2cm}\includegraphics[width=3.7in,height=2.5in]{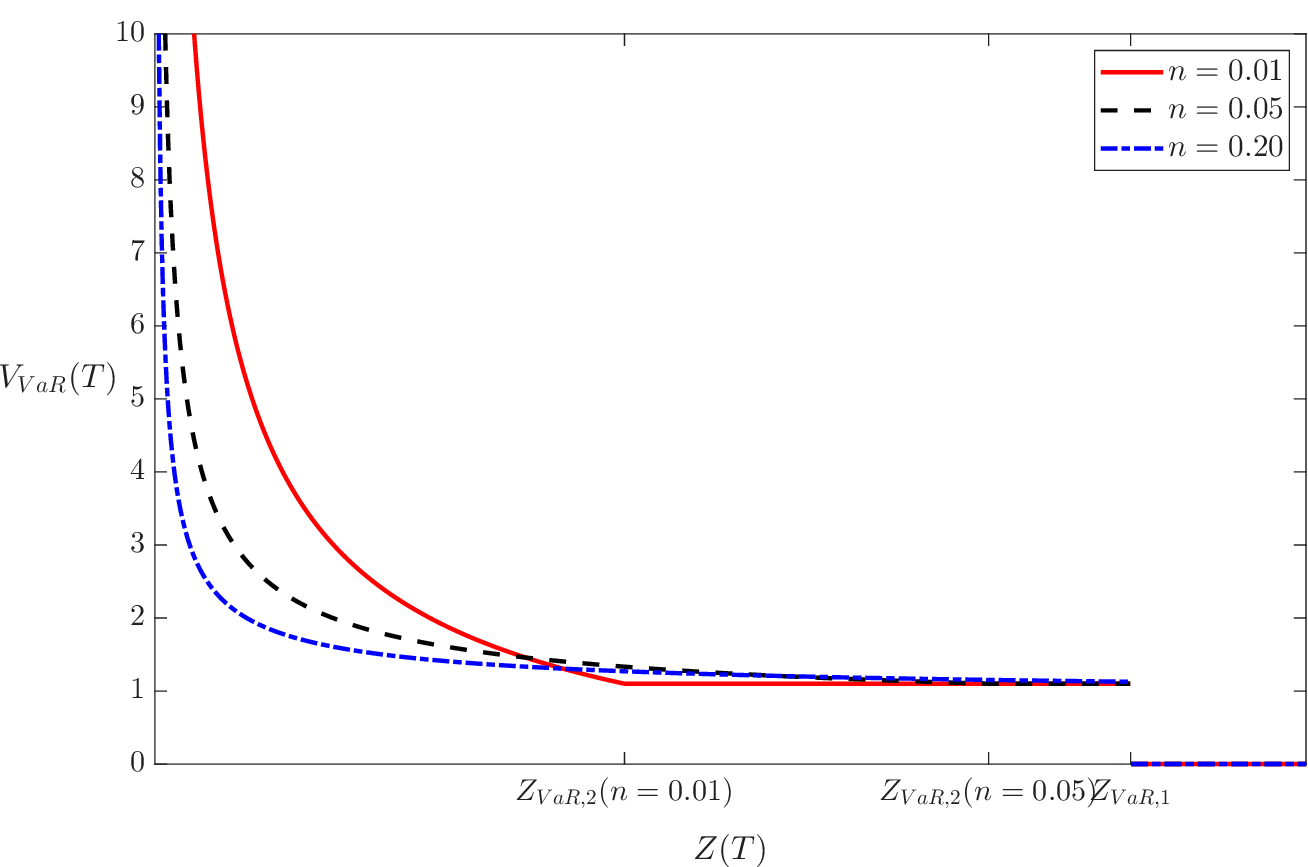}
        \caption{Sensitivity with $n$}
        \label{fig:sensi_n}
    \end{subfigure}\\

    \begin{subfigure}[b]{0.5\textwidth}
    \hspace*{-1.2cm}\includegraphics[width=3.7in,height=2.5in]{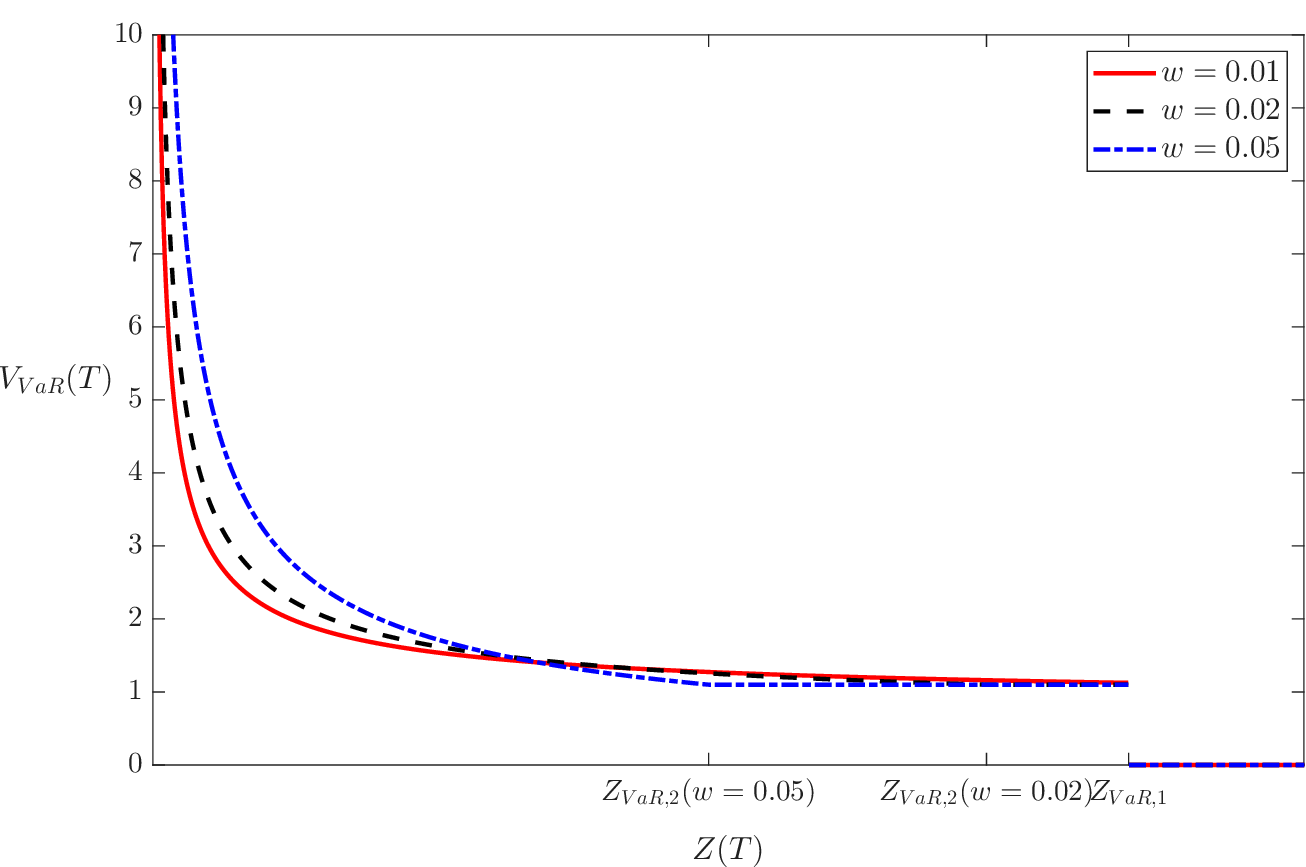}
        \caption{Sensitivity with $w$}
        \label{fig:sensi_w}
    \end{subfigure}
    \caption{Sensitivity analysis over parameters $\delta$, $n$, and $w$}
    \label{sensistivity_anlaysis_1}
\end{figure}

Figure \ref{fig:sensi_delta} studies the effect of $\delta$. A higher value of $\delta$ raises the terminal payoff on the curved branch and shifts the cutoff point for the flat segment, $Z_{\text{VaR},2}$ to the right. Therefore, as $\delta$ becomes larger, the firm value takes the floor $\underline{V}$ over fewer states. Intuitively, when effort is more effective in increasing the firm value, the manager can obtain a higher payoff in 
all states with a positive firm value, and hence, the value is above the VaR floor in a wider range of states, reducing the importance of the VaR protection. For sufficiently large $\delta$, the flat segment disappears and the payoff switches from regime (d) to (e).

Figure \ref{fig:sensi_n} visualizes the firm value as the option-based compensation parameter $n$ varies. As $n$ increases, the curved branch of the firm value shifts downwards in good states, and the cutoff $Z_{\text{VaR},2}$
moves to the right. The firm value takes the form of Case (d) for small $n$, and takes the form of Case (e) for large $n$. Hence, the constraint at $\underline{V}$ is effective over a narrower range of intermediate states. Economically, a larger $n$ makes the manager's compensation more sensitive to the firm's performance above the strike price, as one unit of firm value above the strike translates into more compensation. Because the manager's utility is concave, the marginal utility generated by an additional unit of firm value is lower in good states than in intermediate states. A higher $n$ enlarges this difference even more, making it attractive to give up some upside firm value in good states and reallocate value toward intermediate states.

Figure \ref{fig:sensi_w} studies the effect of the fixed salary $w$. A higher fixed salary raises the payoff in favorable states but also lowers the cutoff $Z_{\text{VaR},2}$, so the VaR constraint is effective for a wider range of states. Intuitively, a higher fixed salary increases the manager's guaranteed compensation independently of the firm's performance, thereby reducing the marginal utility gain from the option component across all states. As a result, in some intermediate states, instead of preserving a firm value slightly above $\underline{V}$, the manager chooses to maintain only the lowest allowed value, the regulatory floor $\underline{V}$. At the same time, the region with a payoff larger than $\underline{V}$ is concentrated in a smaller set of sufficiently favorable states. Given the budget constraint, the manager tends to reallocate terminal value toward those good states, where the option component is more effective at generating utility.

The results from the sensitivity analysis on $n$ and $w$ suggest that the firm should carefully design the manager's compensation structure depending on the company's risk appetite. Given a small $n$ and a large $w$, the manager is more willing to take risky actions. The manager has fewer incentives to smooth firm value in unfavorable states. That is, the manager generates higher firm value in good states, and firm value drops rapidly towards the VaR floor, $\underline{V}$, as the state worsens. In contrast, when $n$ is large and $w$ is small, the manager becomes more cautious and responsible, and the firm value distribution is smoother. We observe a higher firm value in intermediate states at the cost of sacrificing some upside payoff in good states. The intuition is that a larger option component and a lower fixed salary make the manager’s compensation more sensitive to firm performance. As a result, the manager has stronger incentives to preserve firm value in intermediate states rather than concentrate payoff only in the most favorable states.

\subsection{Optimal time-t firm value and project choice}
The following proposition presents explicit expressions for the firm's time-t value and project choice.
We define 
\begin{equation*}
	h(t)=\left\{ 
	\begin{array}{ll}
\frac{\delta ^2}{\theta ^2 -2r}(e^{(\theta ^2 -2r)T} - e^{(\theta ^2 -2r)t}) & ~ \theta ^2  \neq 2r ,\\
\delta ^2 (T-t) & ~ \theta ^2  = 2r.
  \end{array}
	\right.
\end{equation*}
\begin{prop}\label{prop:CRRA_t}
    With the CRRA utility, we have the following assertions.\\
    For the benchmark manager:
    \begin{enumerate}
        \item The optimal time-$t$ firm value is
    \[
    V_0(t)=\left(K-\frac{w}{n}\right)e^{-r(T-t)}N(d_2(Z_0)) + n^{\frac{1-\gamma}{\gamma}}\frac{e^{\Gamma(t)}}{(\lambda_0Z(t))^{\frac{1}{\gamma}}}N(d_1(Z_0))-Z(t)\lambda_0e^{-(\theta^2-2r)t}h(t),
    \]
    where $Z_0,\lambda_0$ are as in Proposition \ref{prop:non-risk}, and
    \begin{align*}
            d_2(x)=&\dfrac{\log\frac{x}{Z(t)}+(r-\frac{1}{2}\theta^2)(T-t)}{\theta\sqrt{(T-t)}},\\
            d_1(x)=&d_2(x)+\frac{1}{\gamma}\theta\sqrt{T-t},\\
            \Gamma(t)=&\frac{1-\gamma}{\gamma}\left(r+\frac{\theta^2}{2\gamma}\right)(T-t),
        \end{align*}
        $N(\cdot)$ is the standard normal distribution function.
        \item The optimal project choice at time $t$ is
        \begin{align*}
            v_0(t)
        =&\left(K-\frac{w}{n}\right)e^{-r(T-t)}\frac{\phi(d_2(Z_0))}{\sqrt{T-t}}+\frac{ n^{\frac{1-\gamma}{\gamma}}e^{\Gamma(t)}}{(\lambda_0 Z(t))^{\frac{1}{\gamma}}}\left(\frac{\theta}{\gamma}N(d_1(Z_0))+\frac{\phi(d_1(Z_0))}{\sqrt{T-t}}\right)\\
        &+\theta Z(t)e^{-(\theta^2-2r)t}\lambda_0h(t),
        \end{align*}
    where $\phi(\cdot)$ is the probability density function of a standard normal random variable.
    \end{enumerate}

    \begin{fleqn}[0pt]
    For the VaR manager:
    \begin{enumerate}
        \item The optimal time-$t$ firm value is given by
        \begin{align*}
            &V_{\text{VaR}}(t)\\
            &=A_1\left(K-\frac{w}{n}\right)e^{-r(T-t)}+A_2\frac{n^{\frac{1-\gamma}{\gamma}}e^{\Gamma(t)}}{(\lambda_{\text{VaR}} Z(t))^{\frac{1}{\gamma}}}+A_3\underline{V}e^{-r(T-t)}-Z(t)\lambda_{\text{VaR}}e^{-(\theta^2-2r)t}h(t).
        \end{align*}

        If $\underline{V}<K$,
    
    \begin{align*} A_1=& N(d_2(\min(Z_{\text{VaR},1},Z_{\text{VaR}}(\beta_2))))+\left(N(d_2(Z_{\text{VaR}}(\beta_1)))-N(d_2(Z_{\text{VaR},1}))\right)^+,\\
    A_2=& N(d_1(\min(Z_{\text{VaR},1},Z_{\text{VaR}}(\beta_2))))+\left(N(d_1(Z_{\text{VaR}}(\beta_1)))-N(d_1(Z_{\text{VaR},1}))\right)^+,\\
    A_3=& \left(N(d_2(Z_{\text{VaR},1}))-N(d_2(Z_{\text{VaR}}(\beta_2)))\right)^+.
    \end{align*}

        If $\underline{V}\ge K$,
        \begin{align*}
            A_1=&N(d_2(\min(Z_{\text{VaR},1},Z_{\text{VaR},2})))+\left(N(d_2(Z_{\text{VaR}}(\beta_1)))-N(d_2(Z_{\text{VaR},1}))\right)^+,\\
            A_2=&N(d_1(\min(Z_{\text{VaR},1},Z_{\text{VaR},2})))+\left(N(d_1(Z_{\text{VaR}}(\beta_1)))-N(d_1(Z_{\text{VaR},1}))\right)^+,\\
            A_3=& \left(N(d_2(Z_{\text{VaR},1}))-N(d_2(Z_{\text{VaR},2}))\right)^+.
        \end{align*}

     \item The optimal project choice at time $t$ is given by
     \begin{align*}
         v_{\text{VaR}}(t)
         =&B_1\frac{(K-\frac{w}{n})e^{-r(T-t)}}{\sqrt{T-t}}+\left(\frac{\theta}{\gamma}B_2+\frac{1}{\sqrt{T-t}}B_3\right)\frac{n^{\frac{1-\gamma}{\gamma}}e^{\Gamma(t)}}{\lambda_{\text{VaR}}^{\frac{1}{\gamma}} Z(t)^{\frac{1}{\gamma}}}\\
         &+B_4\frac{\underline{V}e^{-r(T-t)}}{\sqrt{T-t}}+\theta Z(t)e^{-(\theta^2-2r)t}\lambda_{\text{VaR}}h(t).
     \end{align*}

     If $\underline{V}<K$,
     \begin{align*}
         B_1=&\phi(d_2(\min(Z_{\text{VaR},1},Z_{\text{VaR}}(\beta_2))))+\left(\phi(d_2(Z_{\text{VaR}}(\beta_1)))-\phi(d_2(Z_{\text{VaR},1}))\right)\mathbf{1}_{\{Z_{\text{VaR}(\beta_1)}>Z_{\text{VaR},1}\}},\\
         B_2=&N(d_1(\min(Z_{\text{VaR},1},Z_{\text{VaR}}(\beta_2))))+\left(N(d_1(Z_{\text{VaR}}(\beta_1)))-N(d_1(Z_{\text{VaR},1}))\right)^+,\\
         B_3=&\phi(d_1(\min(Z_{\text{VaR},1},Z_{\text{VaR}}(\beta_2))))+\left(\phi(d_1(Z_{\text{VaR}}(\beta_1)))-\phi(d_1(Z_{\text{VaR},1})) \right)\mathbf{1}_{\{Z_{\text{VaR}}(\beta_1)>Z_{\text{VaR},1}\}},\\
         B_4=&\left(\phi(d_2(Z_{\text{VaR},1}))-\phi(d_2({Z}_{\text{VaR}}(\beta_2)))\right)\mathbf{1}_{\{Z_{\text{VaR},1}>Z_{\text{VaR}}(\beta_2)\}}.
     \end{align*}

     If $\underline{V}\ge K$,
     \begin{align*}
         B_1=&\phi(d_2(\min(Z_{\text{VaR},1},Z_{\text{VaR},2})))+\left(\phi(d_2(Z_{\text{VaR}}(\beta_1)))-\phi(d_2(Z_{\text{VaR},1}))\right)\mathbf{1}_{\{Z_{\text{VaR}(\beta_1)}>Z_{\text{VaR},1}\}},\\
         B_2=&N(d_1(\min(Z_{\text{VaR},1},Z_{\text{VaR},2})))+\left(N(d_1(Z_{\text{VaR}}(\beta_1)))-N(d_1(Z_{\text{VaR},1}))\right)^+,\\
         B_3=&\phi(d_1(\min(Z_{\text{VaR},1},Z_{\text{VaR},2})))+(\phi(d_1(Z_{\text{VaR}}(\beta_1)))-\phi(d_1(Z_{\text{VaR},1})) )\mathbf{1}_{\{Z_{\text{VaR}}(\beta_1)>Z_{\text{VaR},1}\}},\\
         B_4=&\left(\phi(d_2(Z_{\text{VaR},1}))-\phi(d_2(Z_{\text{VaR},2}))\right)\mathbf{1}_{\{Z_{\text{VaR},1}>Z_{\text{VaR},2}\}}.
     \end{align*}
     
    \end{enumerate}
    \end{fleqn}
\end{prop}

\begin{figure}[!htbp]
    \centering
    
    \begin{subfigure}[b]{0.48\textwidth}
        \centering
    \includegraphics[width=3.2in,height=2.5in]{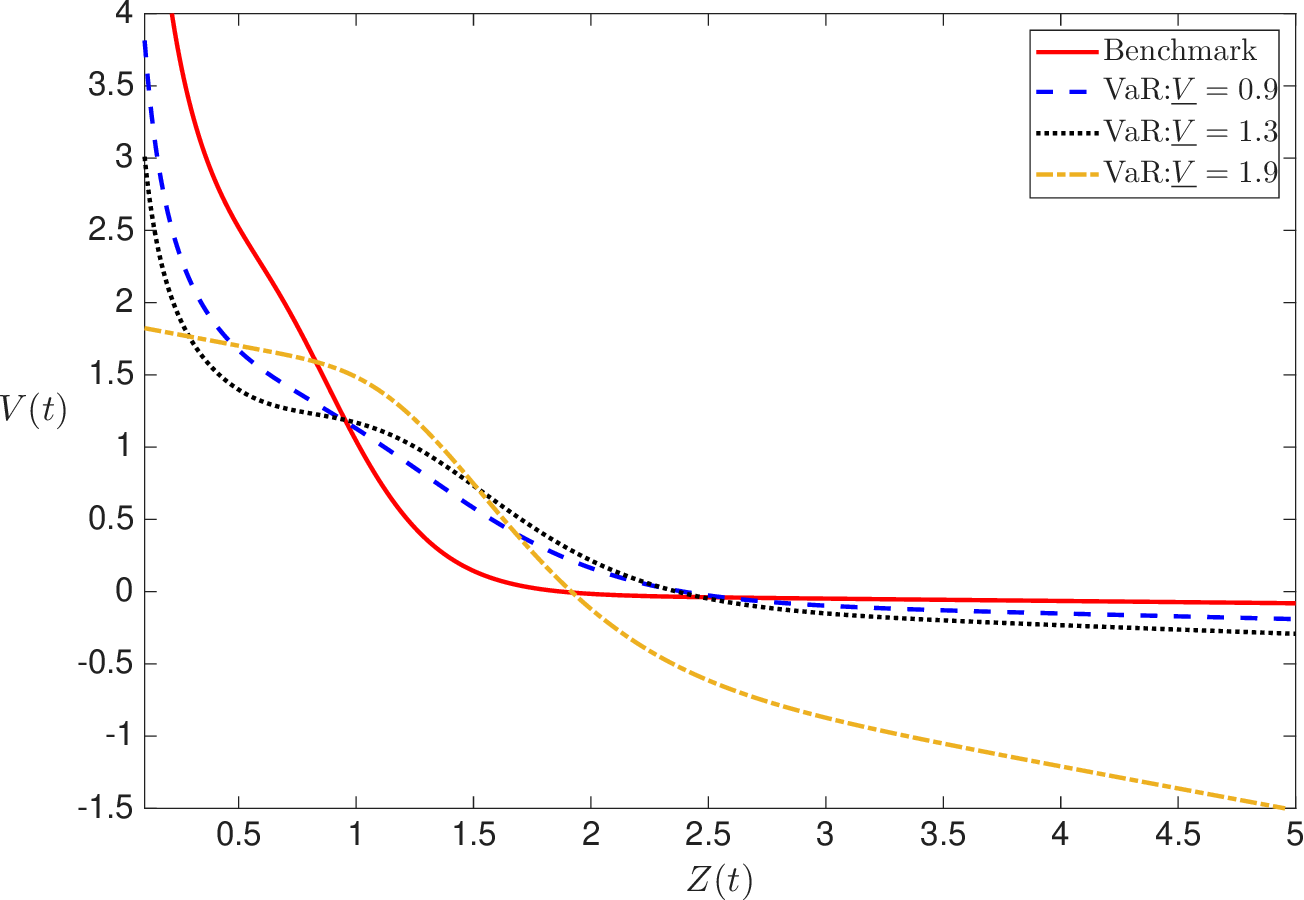}
        \caption{Firm value under $\alpha=0.05$}
        \label{fig:Vt_V}
    \end{subfigure}
    \hfill
    \begin{subfigure}[b]{0.48\textwidth}
        \centering
        \includegraphics[width=3.2in,height=2.5in]{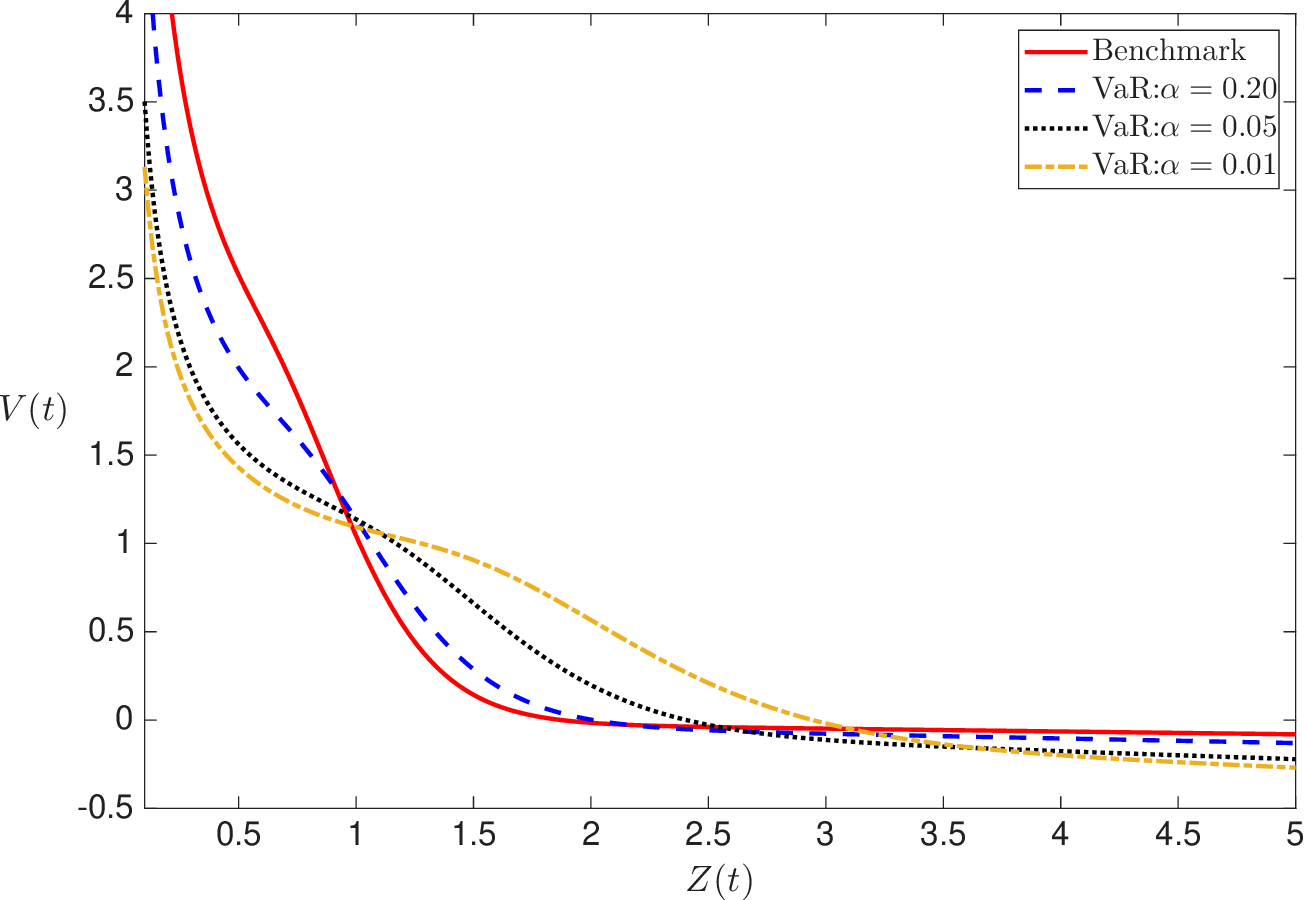}
        \caption{Firm value under $\underline{V}=1.1$}
        \label{fig:Vt_alpha}
    \end{subfigure}

    \vspace{0.5em}

    \begin{subfigure}[b]{0.48\textwidth}
        \centering
        \includegraphics[width=3.2in,height=2.5in]{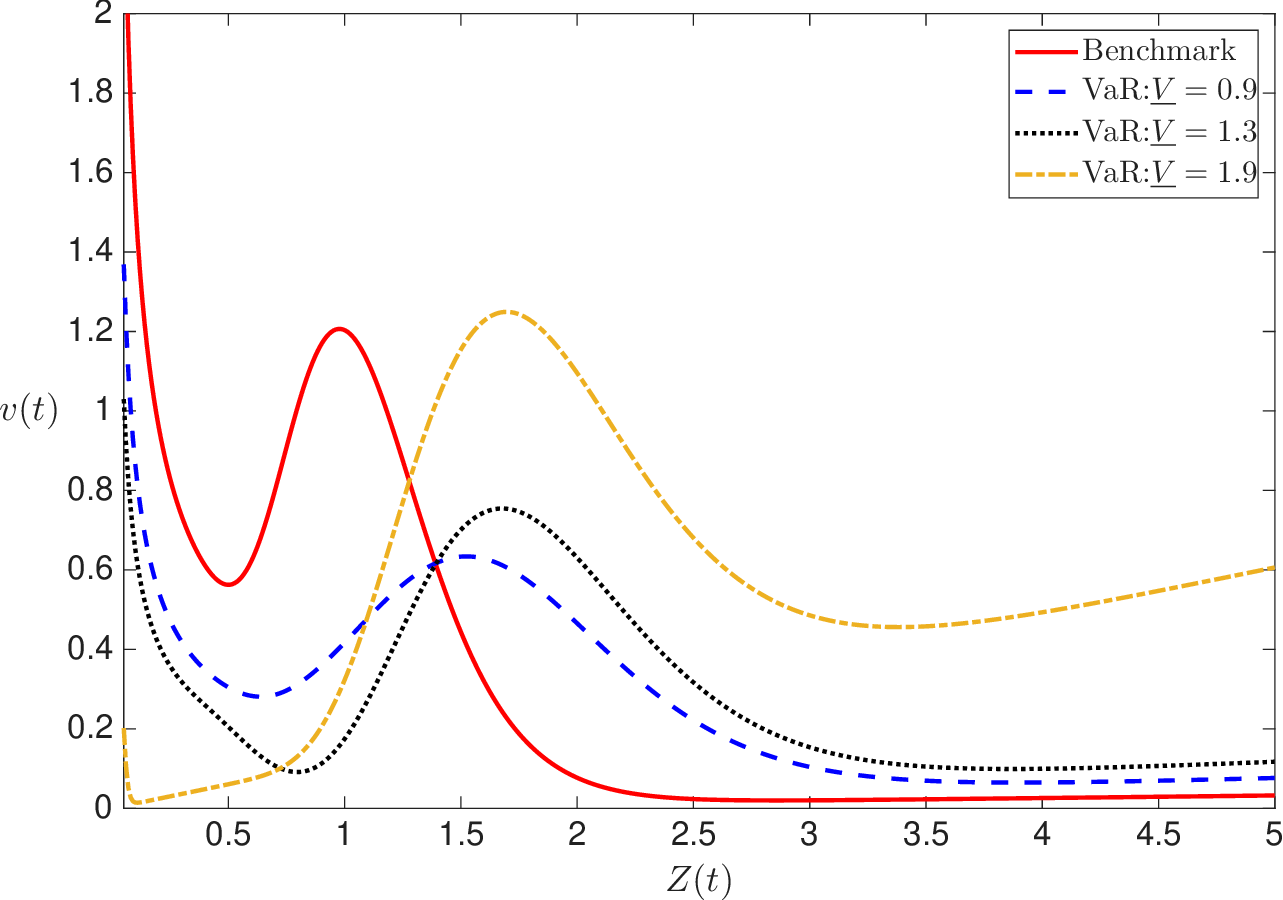}
        \caption{Volatility under $\alpha=0.05$}
        \label{fig:project_t_V}
    \end{subfigure}
    \hfill
    \begin{subfigure}[b]{0.48\textwidth}
        \centering
        \includegraphics[width=3.2in,height=2.5in]{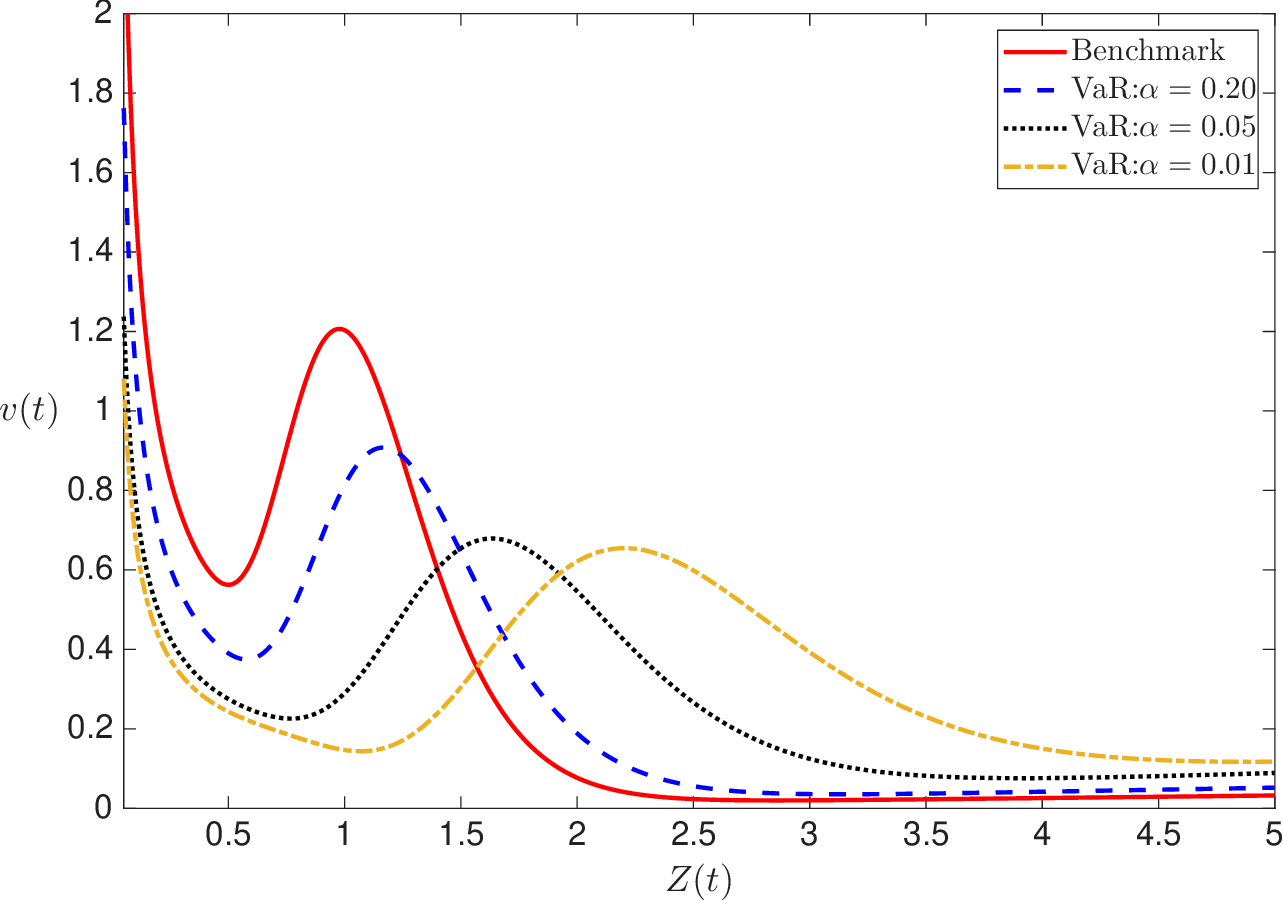}
        \caption{Volatility under $\underline{V}=1.1$}
        \label{fig:project_t_alpha}
    \end{subfigure}

    \caption{Sensitivity analysis of the firm value $V(t)$ and volatility $v(t)$ at $t=0.5$}
    \label{fig:sensitivity_analysis_2}
\end{figure}

Figures \ref{fig:Vt_V} and \ref{fig:Vt_alpha} visualize the time-$t$ firm value with different VaR floors $\underline{V}$ and tail probabilities $\alpha$. We select an intermediate time point $t=0.5$ for illustration. We consider three VaR floors $\underline{V}=0.9, 1.3, 1.9$, corresponding to the three regimes of $\underline{V}$. The tail probability $\alpha$ is set to $0.01, 0.05, 0.2$, representing strict, moderate, and loose VaR constraints. The time-$t$ firm value shown in Figures \ref{fig:Vt_V} and \ref{fig:Vt_alpha} exhibits a similar pattern to the terminal firm value: compared with the benchmark scenario, the VaR manager gives up some high firm value in good states, and obtains a higher value in intermediate states in order to satisfy the VaR constraint. However, in sufficiently bad states, a stricter VaR constraint can lower the time-$t$ firm value. Since the VaR requirement is imposed only on the terminal value $V(T)$, the manager needs to sacrifice interim firm value in adverse states in order to preserve the required downside protection at maturity. Since the firm is not required to remain nonnegative before $T$, the time-$t$ value in bad states may drop below $0$.

Figures \ref{fig:project_t_V} and \ref{fig:project_t_alpha} plot the optimal time-$t$ project volatility choices. A hump-shaped project volatility is observed for both the benchmark and the VaR manager. When the state is good, the manager expects a terminal firm value well above the strike price, so additional volatility may introduce undesirable risk for the risk-averse manager. In the intermediate states, however, the option component of compensation is most sensitive to changes in firm value: higher volatility increases the chance of making the option deep in the money, while the downside is partially protected since the manager is guaranteed to receive the fixed salary and the level of firm value below the strike does not affect the manager's compensation. As a result, the manager chooses higher project volatility in this region. When the state deteriorates further, the probability that the option finishes in the money becomes small, so the option payoff benefits less from high volatility, and the manager again reduces project risk.

The value of $(\underline{V},\alpha)$ affects the project choices of the VaR manager. A tighter VaR constraint lowers project volatility in good and intermediate states, because the manager must make the distribution of terminal firm value less likely to fall below the floor. Consequently, the region in which risk-taking is attractive is pushed toward worse states, so the hump in $v(t)$ moves to the right as the constraint becomes stricter. Moreover, when the VaR constraint is sufficiently strict, the manager may engage in `gambling for recovery' behavior in very adverse states. The key reason is that in our model, a higher project volatility not only increases risk but also raises the expected growth rate of firm value through the positive risk-premium parameter $\theta>0$. Under a tight VaR requirement, the manager must improve outcomes even in relatively bad states to satisfy the downside constraint at maturity. This makes riskier projects attractive in adverse states since they provide both greater upside potential and a higher expected return.

\section{Conclusion}\label{sec:conclusion}

This paper studies a dynamic corporate risk management problem in which a risk-averse manager chooses effort and project volatility. We assume the manager is compensated with a fixed salary and an option-based incentive, resulting in a non-concave objective for the firm's value. Using the martingale method, concavification, and quantile formulation techniques, we derive explicit expressions for the optimal terminal firm value, effort level, and project choice for a manager subject to VaR constraints.

Our study shows that the VaR constraint substantially reshapes the firm value structure and managerial decisions. Depending on the VaR floor and tail probability, the firm's value can be binary, ternary, or quaternary. We further show that a binding VaR constraint induces higher managerial effort. In addition, compared to the benchmark case, the VaR constraint reduces the bankruptcy probability in most scenarios; however, if the VaR floor is sufficiently high and the tail probability is low, the bankruptcy probability may increase. A sensitivity analysis shows that higher productivity of managerial effort leads to a higher terminal firm value in all non-bankruptcy states. Moreover, a high fixed salary and a lower option-based incentive encourage the manager to make risk-seeking decisions. The manager becomes less responsible and has fewer incentives to smooth firm value across states. 

We suggest that the firm should hire a skilled manager and design the compensation package based on the firm's risk appetite and shareholders' preferences. Our results also indicate that the VaR constraint should be set carefully: setting the floor or tail probability too aggressively may distort managerial risk-taking, increase the default probability, and incentivize gambling behavior in adverse states. In contrast, a moderate VaR requirement can provide downside protection and reduce the default probability.

\begin{appendices}
\renewcommand{\theequation}{\thesection.\arabic{equation}}
\setcounter{equation}{0}
\renewcommand{\thefigure}{\thesection.\arabic{figure}}
\setcounter{figure}{0}
\section{Proofs}\label{appendix:proof}

\subsection{Proof of Proposition 2.1}\label{appendix_1}

By Ito's formula, we have
\begin{align*}
dZ(t)=&-rZ(t)dt-\theta Z(t)dW(t),\\
d(Z(t)V(t))=&Z(t)dV(t)+V(t)dZ(t)+d\langle Z, V\rangle_t\\
=&\delta Z(t)u(t)dt+Z(t)(v(t)-\theta V(t))dW(t).
\end{align*}
Integrating from $t$ to $T$, we get
\[
Z(T)V(T)-Z(t)V(t)=\delta\int_t^T Z(s)u(s)ds+\int_t^T Z(s)(v(s)-\theta V(s))dW(s).
\]
By integrability assumptions, we have
\[
Z(t)V(t)=\mathbb{E}\left[Z(T)V(T)-\delta\int_t^T Z(s)u(s)ds |\mathcal{F}_t \right].
\]
Since $Z(t)>0$, we obtain
\[
V(t)=\frac{1}{Z(t)}\mathbb{E}
\left[Z(T)V(T)-\delta\int_t^T Z(s)u(s)ds|\mathcal{F}_t\right].
\]
\subsection{Proof of Proposition 3.3}\label{appendix_2}

Let $\mathbb{G} := \{F^{-1}(\cdot)(t) : F( \cdot ) \in  \mathbb{F} \}$ be the corresponding set of quantile functions, or
\begin{equation*}
\begin{aligned}
\mathbb{G} = \{G(\cdot) : [0, 1) \to \mathbb{R}^{+},\text{ nondecreasing, right continuous with left limits (RCLL)} \},
\end{aligned}
\end{equation*}
where $G(1) := G(1-)$.

We employ the quantile formulation to solve (4). The quantile formulation, developed in a series of papers including \cite{schied2004neyman,carlier2006law,jin2008behavioral,he2011portfolio,carlier2011optimal}, is a technique
for solving optimal terminal payoffs in portfolio choice problems. This technique can be applied when
the objective function and all constraints, except for the initial budget constraint, are law-invariant, and the objective function is increasing in terminal wealth. The key idea is to derive the optimal quantile function of the terminal payoff in terms of the state variables.
In our setting, it is obvious that the objective function is increasing in terminal firm value. The optimal solution can be obtained with the help of the following lemma from \cite{jin2008behavioral},

\begin{lemma}[\cite{jin2008behavioral}]\label{lemma:quantile}

For any lower-bounded random variable $V$ with quantile function $G$, we have
\[
\E[Z(T) G(1-F_{Z}(Z(T)))] \le \E[Z(T) V].
\]
Furthermore, if $\E [Z(T) G(1-F_{Z}(Z(T)))] < \infty$, then the inequality
becomes equality if and only if 
\[
V = G(1-F_{Z}(Z(T))), ~a.s.
\]
\end{lemma}

We consider the quantile formulation of (4),
\begin{equation}\label{prob:QF}
\begin{aligned}
 \max _{G(\cdot) \in \mathbb{G}, u} ~ & J(G(\cdot),u) = \int _0^1 U(w + n(G(s) - K)^{+})ds - \E \left[\frac{1}{2}\int _0^T u^2(t)dt\right]\\
\text{subject to} ~ & \int _0^1 F_{Z} ^{-1}(1-s)G(s)ds - \E \left[\delta \int _0^T Z (t) u(t)dt\right] = V(0),\\
&G(\alpha) \ge \underline{V}, ~G(0) \ge 0.
\end{aligned}
\end{equation}

The proof of Proposition 3.3 is decomposed into four steps. First, we apply the Lagrange dual method to solve \eqref{prob:QF} with the objective function replaced by its concave envelope. Second, using the optimizer from the first step, we determine whether any part of the solution lies in the region where the concave envelope differs from the original objective function; whenever this occurs, we re-optimize over that region under the original objective. Third, we prove the existence of Lagrange multipliers. Finally, calculate the optimal project choice at time $t$.

\noindent \textit{Step 1. Express $G(\cdot)$ and $u(\cdot)$ in terms of the Lagrange multiplier and solve the optimization problem on the concave envelope.}

Define
\begin{equation}\label{eq:lag}
\begin{aligned}
J_{\lambda }(G(\cdot), u(\cdot))=&\int _0^1 U(w + n(G(s) - K)^{+})ds-\mathbb{E}\left[\dfrac{1}{2}\int_0^Tu^2(t)dt\right] \\
&- \lambda\left( \int _0^1 F_{Z} ^{-1}(1-s)G(s)ds-\mathbb{E}\left[\delta\int_0^T Z(t)u(t)dt\right]\right), ~\lambda>0,
\end{aligned}
\end{equation}
and consider
\begin{equation}
\begin{aligned}
 \max _{G(\cdot) \in \mathbb{G},u(\cdot)} ~ & J_{\lambda }(G(\cdot),u(\cdot)) \\
\text{subject to} ~ &G(\alpha) \ge \underline{V}, ~G(0) \ge 0.
\end{aligned}
\end{equation}
Define 
\begin{align*}
    J_{1}(G(\cdot), \lambda)&:=\int _0^1 U(w + n(G(s)-K)^{+}) - \lambda F_{Z} ^{-1}(1-s)G(s)ds,\\
    J_2(u(\cdot), \lambda)&:=\mathbb{E}\left[\int_0^T\lambda\delta Z(t)u(t)-\dfrac{1}2{u^2(t)dt}\right],
\end{align*}
and hence $J_{\lambda }(G(\cdot),u(\cdot))=J_1(G(\cdot), \lambda)+J_2(u(\cdot), \lambda)$. We can optimize $J_1(G(\cdot), \lambda)$ and $J_2(u(\cdot), \lambda)$ separately. 

The integrand of $J_2(u(\cdot), \lambda)$ is a quadratic function in $u$, and hence the optimal $u_\lambda(t)=\lambda\delta Z(t)$. To optimize $J_1(G(\cdot),\lambda)$, since the manager's objective function $U(w+n(G(s)-K)^+)$ is not concave in the variable $G(s)$, we first solve the problem with the concavified objective function:
\begin{equation}\label{prob:QFlag}
\begin{aligned}
\max_{G(\cdot)\in
\mathbb{G}} &\mathcal{L}(G(\cdot),\lambda):= U_1(G(s)) - \lambda F_{Z} ^{-1}(1-s)G(s)\\
\text{subject to} ~ &G(\alpha) \ge \underline{V}, ~G(0) \ge 0,
\end{aligned}
\end{equation}
where
\begin{equation*}
\mathcal{L}(G(\cdot),\lambda)=
\begin{cases}
    U(w)+(n\beta_1-\lambda F_Z^{-1}(1-s))G(s),&G(s)\leq \xi_1,\\
    U(w+n(G(s)-K))-\lambda F_Z^{-1}(1-s)G(s), &G(s)>\xi_1.  
\end{cases}
\end{equation*}
 We have $\alpha_2(\lambda)\le\alpha_1(\lambda,\beta_1)$ if $\underline{V}\le\xi_1$ and $\alpha_2(\lambda)>\alpha_1(\lambda,\beta_1)$ if $\underline{V}>\xi_1$.
There are five cases:
\begin{enumerate}
\item $\underline{V} \le \xi_1, \alpha < \alpha_1(\lambda,\beta_1)$. The maximizer of \eqref{prob:QFlag} is given by 
\begin{equation}\label{eq:G1}
	G _{\lambda , \underline{V},\alpha}(s):=\left\{ 
	\begin{array}{ll}
K + \frac{1}{n} ( I(\frac{1}{n} \lambda F_{Z} ^{-1}(1- s)) - w) & ~ \alpha_1(\lambda,\beta_1) \le  s \le 1,\\
\underline{V} & ~ \alpha \le  s < \alpha_1(\lambda,\beta_1),\\
		0 & ~ 0 \le s < \alpha ,
  \end{array}
	\right.
\end{equation}

\item $\underline{V} \le \xi_1, \alpha \ge \alpha_1(\lambda,\beta_1)$.  The maximizer of \eqref{prob:QFlag} is given by 
\begin{equation}\label{eq:G2}
	G _{\lambda ,\underline{V},\alpha}(s):=\left\{ 
	\begin{array}{ll}
K + \frac{1}{n} ( I(\frac{1}{n} \lambda F_{Z} ^{-1}(1- s)) - w) & ~ \alpha_1(\lambda,\beta_1) \le  s \le 1,\\
		0 & ~ 0 \le s < \alpha_1(\lambda,\beta_1) ,
  \end{array}
	\right.
\end{equation}


\item $\underline{V} > \xi_1, \alpha < \alpha_1(\lambda,\beta_1)<\alpha_2(\lambda)$. The maximizer of \eqref{prob:QFlag} is given by 
\begin{equation}\label{eq:G3}
	G _{\lambda ,\underline{V},\alpha}(s):=\left\{ 
	\begin{array}{ll}
K + \frac{1}{n} ( I(\frac{1}{n} \lambda F_{Z} ^{-1}(1- s)) - w) & ~ \alpha_2(\lambda) \le  s \le 1,\\
\underline{V} & ~ \alpha \le  s < \alpha_2(\lambda),\\
		0 & ~ 0 \le s < \alpha ,
  \end{array}
	\right.
\end{equation}

\item $\underline{V} > \xi_1,  \alpha_1(\lambda,\beta_1)\leq \alpha < \alpha_2(\lambda)$.  The maximizer of \eqref{prob:QFlag} is given by \begin{equation}\label{eq:G4}
	G_{\lambda ,\underline{V},\alpha}(s)=\left\{ 
	\begin{array}{ll}
	K + \frac{1}{n} ( I(\frac{1}{n} \lambda F_{Z} ^{-1}(1- s)) - w) & ~ \alpha_2(\lambda) \le s \le 1 ,\\
	\underline{V} & ~ \alpha \le  s < \alpha_2(\lambda),\\
K + \frac{1}{n} ( I(\frac{1}{n} \lambda F_{Z} ^{-1}(1- s)) - w) & ~ \alpha_1(\lambda,\beta_1) \le s < \alpha, \\
0 & ~ 0 \le s < \alpha_1(\lambda,\beta_1).
  \end{array}
	\right.
\end{equation}

\item $\underline{V} > \xi_1,  \alpha_1(\lambda,\beta_1) < \alpha_2(\lambda)\leq \alpha$.  The maximizer of \eqref{prob:QFlag} is given by 
\begin{equation}\label{eq:G5}
	G _{\lambda ,\underline{V},\alpha}(s):=\left\{ 
	\begin{array}{ll}
K + \frac{1}{n} ( I(\frac{1}{n} \lambda F_{Z} ^{-1}(1- s)) - w) & ~ \alpha_1(\lambda,\beta_1) \le  s \le 1,\\
		0 & ~ 0 \le s < \alpha_1(\lambda,\beta_1) ,
  \end{array}
	\right.
\end{equation}
\end{enumerate}

By the definition of the concave envelope in (5), the concave envelope and the original objective function disagree on the interval $(0,\xi_1)$. Note that in Case 1, the optimal terminal firm value $G_{\lambda,\underline{V},\alpha}(s)=\underline{V}\in(0,\xi_1)$ for $\alpha\le s <\alpha_1(\lambda,\beta_1)$. In other cases, the optimal terminal value always takes on values where the two functions are equal. Therefore, we consider the interval $s\in[\alpha,\alpha_1(\lambda,\beta_1))$ separately with the original objective function.
We further decompose the case $\underline{V} \le \xi_1, \alpha < \alpha_1(\lambda,\beta_1)$ into two sub-cases.

\noindent \textit{Step 2. Solve the optimization problem in the region where the original objective and the concave envelope disagree.}

\begin{enumerate}[label=1.\arabic*]
\item $K \le \underline{V} \le\xi_1$, $\alpha<\alpha_1(\lambda,\beta_1)$. The original objective $U_0(x)$ is concave on $[\underline{V},\xi_1]$, and we obtain the optimal solution by comparing the local maximum and the boundary point. The optimal solution for $\alpha \le s < \alpha_1(\lambda,\beta_1)$ is divided into two cases:
\begin{enumerate}
    \item For $\alpha<\alpha_2(\lambda)$,
\begin{equation*}
    G_{\lambda,\underline{V},\alpha}(s)= \begin{cases}
        K + \frac{1}{n} \left( I\left(\frac{1}{n} \lambda F_{Z} ^{-1}(1- s)\right) - w\right), &\alpha_2(\lambda)\le s < \alpha_1(\lambda,\beta_1),\\
        \underline{V}, &\alpha\le s< \alpha_2(\lambda).
    \end{cases}
\end{equation*}
\item For $\alpha_2(\lambda)\le \alpha<\alpha_1(\lambda,\beta_1)$,
\begin{equation*}
    G_{\lambda,\underline{V},\alpha}(s)=
        K + \frac{1}{n} \left( I\left(\frac{1}{n} \lambda F_{Z} ^{-1}(1- s)\right) - w\right), \text{ } \alpha\le s <\alpha_1(\lambda,\beta_1).
\end{equation*}
\end{enumerate}

\item $\underline{V}<K,\alpha<\alpha_1(\lambda,\beta_1)$. The original objective $U_0(x)$ is not concave on $[\underline{V},\xi_1)$ , and hence we construct an inner concave envelope $U_2(x): [\underline{V},\infty)\rightarrow[U_0(\underline{V}),\infty)$ on this interval. We define
\begin{equation*}
    U_2(x)=\begin{cases}
        U(w+n(x-K)^+), &x\ge \xi_2\\
        U(w)+n\beta_2(x-\underline{V}), &\underline{V}\le x<\xi_2,
    \end{cases}
\end{equation*}
where $\xi_2\in(K,\xi_1)$ is the unique solution to 
\begin{equation*}
U(w)+U^\prime(w+n(\xi_2-K))n(\xi_2-\underline{V})=U(w+n(\xi_2-K)^+),
\end{equation*}
and $
\beta_2=U^\prime(w+n(\xi_2-K)).$ 
Obviously, we have $\beta_2>\beta_1$, and hence $\alpha_1(\lambda,
\beta_2)<\alpha_1(\lambda,\beta_1)$. $U_2(x)$ and $U_0(x)$ disagree on $x\in[\underline{V},\xi_2]$.
Figure \ref{fig:envelope_small} plots $U_0,U_1,U_2$. $U_2$ is tangent to $U_0$ at $\xi_2$.

\begin{figure}
    \centering
\includegraphics[width=0.8\linewidth]{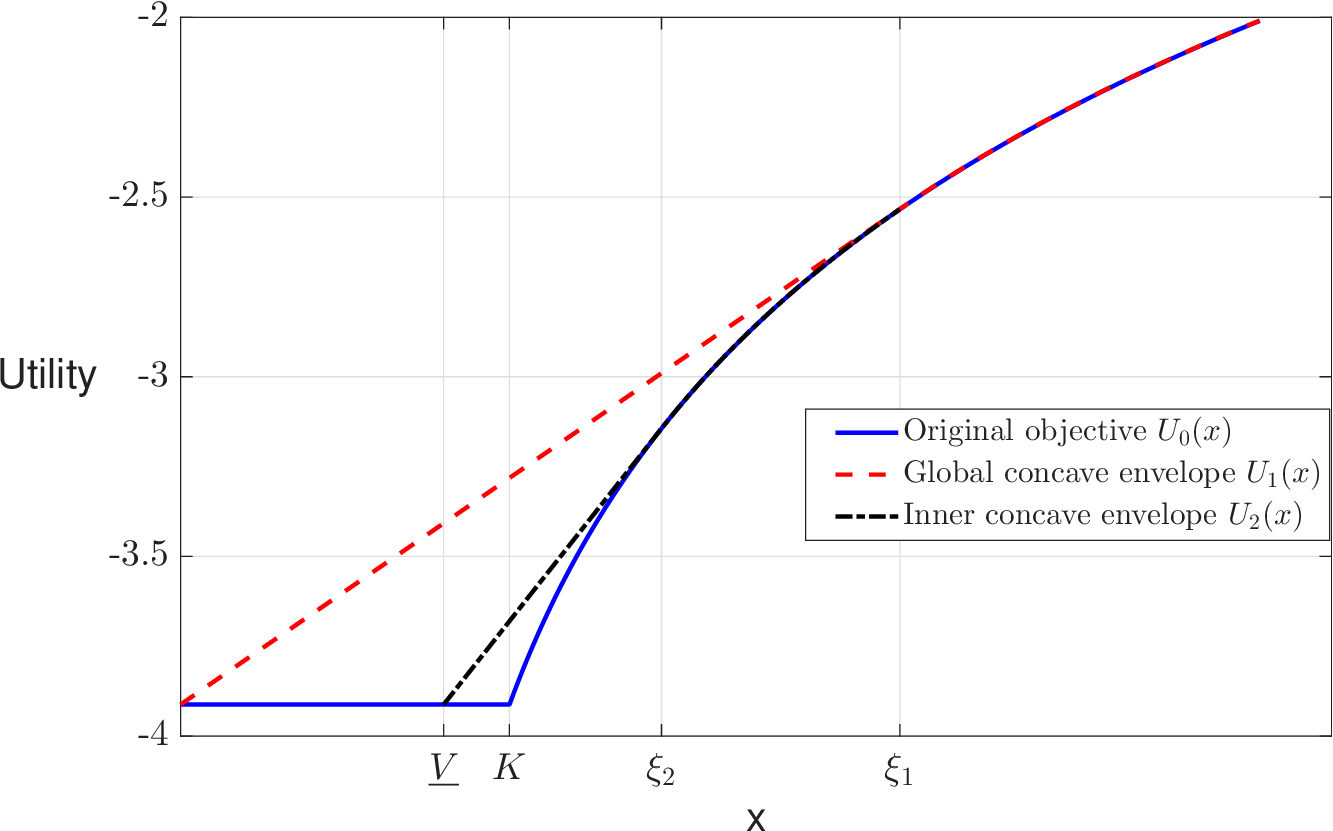}
    \caption{Global and Inner Concave Envelopes}
    \label{fig:envelope_small}
\end{figure}

We consider the optimization problem with the concave envelope on $U_2(x)$. The maximizer is separated into two cases:
\begin{enumerate}
    \item  For $\alpha<\alpha_1(\lambda,\beta_2),$
    \begin{equation*}
    G_{\lambda,\underline{V},\alpha}(s)=\begin{cases}
        K + \frac{1}{n} ( I(\frac{1}{n} \lambda F_{Z} ^{-1}(1- s)) - w), &\alpha_1(\lambda,\beta_2)\le s <\alpha_1(\lambda,\beta_1)\\
        \underline{V}, &\alpha\le s <\alpha_1(\lambda,\beta_2).
    \end{cases}
\end{equation*}
\item For $\alpha_1(\lambda,\beta_2)\le\alpha<\alpha_1(\lambda,\beta_1)$,
\begin{equation*}
    G_{\lambda,\underline{V},\alpha}(s)=
        K + \frac{1}{n} \left( I\left(\frac{1}{n} \lambda F_{Z} ^{-1}(1- s)\right) - w\right), \alpha\le s <\alpha_1(\lambda,\beta_1).
\end{equation*}
\end{enumerate}

Since the optimal terminal firm value takes on values where $U_0(x)$ and $U_2(x)$ are equivalent, the solution is also optimal to the original problem.
\end{enumerate}
In summary, the optimal terminal solution to the original problem (4) is given by the following nine cases.

\begin{enumerate}[label=(\alph*)]
    \item $\underline{V}<K,\alpha<\alpha_1(\lambda,\beta_2)$. The maximizer is given by
    \begin{equation}\label{eq:G1,original}
	G _{\lambda , \underline{V},\alpha}(s):=\left\{ 
	\begin{array}{ll}
K + \frac{1}{n} \left( I\left(\frac{1}{n} \lambda F_{Z} ^{-1}(1- s)\right) - w\right) & ~ \alpha_1(\lambda,\beta_2) \le  s \le 1,\\
\underline{V} & ~ \alpha \le  s < \alpha_1(\lambda,\beta_2),\\
		0 & ~ 0 \le s < \alpha.
  \end{array}
	\right.
\end{equation}
\item $\underline{V}<K,\alpha_1(\lambda,\beta_2)\le \alpha < \alpha_1(\lambda,\beta_1)$ The maximizer is given by
\begin{equation}\label{eq:G2,original}
	G _{\lambda ,\underline{V},\alpha}(s):=\left\{ 
	\begin{array}{ll}
K + \frac{1}{n} \left( I\left(\frac{1}{n} \lambda F_{Z} ^{-1}(1- s)\right) - w\right) & ~ \alpha \le  s \le 1,\\
		0 & ~ 0 \le s < \alpha.
  \end{array}
	\right.
\end{equation}

 \item $\underline{V}<K,\alpha\ge \alpha_1(\lambda,\beta_1)$. The maximizer is given by
\begin{equation}\label{eq:G3,original}
	G _{\lambda ,\underline{V},\alpha}(s):=\left\{ 
	\begin{array}{ll}
K + \frac{1}{n} \left( I\left(\frac{1}{n} \lambda F_{Z} ^{-1}(1- s)\right) - w\right) & ~ \alpha_1(\lambda,\beta_1) \le  s \le 1,\\
		0 & ~ 0 \le s < \alpha_1(\lambda,\beta_1),
  \end{array}
	\right.
\end{equation}
and hence the VaR constraint is inactive.
\item $K\le\underline{V}<\xi_1,\alpha<\alpha_2(\lambda).$ The maximizer is given by 
\begin{equation}\label{eq:G4,original}
	G _{\lambda , \underline{V},\alpha}(s):=\left\{ 
	\begin{array}{ll}
K + \frac{1}{n} \left( I\left(\frac{1}{n} \lambda F_{Z} ^{-1}(1- s)\right) - w\right) & ~ \alpha_2(\lambda) \le  s \le 1,\\
\underline{V} & ~ \alpha \le  s < \alpha_2(\lambda),\\
		0 & ~ 0 \le s < \alpha ,
  \end{array}
	\right.
\end{equation}

\item  $K\le\underline{V}<\xi_1,\alpha_2(\lambda)\le \alpha<\alpha_1(\lambda,\beta_1).$  The maximizer is given by
\begin{equation}\label{eq:G5,original}
	G _{\lambda ,\underline{V},\alpha}(s):=\left\{ 
	\begin{array}{ll}
K + \frac{1}{n} \left( I\left(\frac{1}{n} \lambda F_{Z} ^{-1}(1- s)\right) - w\right) & ~ \alpha \le  s \le 1,\\
		0 & ~ 0 \le s < \alpha.
  \end{array}
	\right.
\end{equation}
\item $K\le\underline{V}<\xi_1, \alpha\ge\alpha_1(\lambda,\beta_1).$ The maximizer is given by 
\begin{equation}\label{eq:G6,original}
	G _{\lambda ,\underline{V},\alpha}(s):=\left\{ 
	\begin{array}{ll}
K + \frac{1}{n} \left( I\left(\frac{1}{n} \lambda F_{Z} ^{-1}(1- s)\right) - w\right) & ~ \alpha_1(\lambda,\beta_1) \le  s \le 1,\\
		0 & ~ 0 \le s < \alpha_1(\lambda,\beta_1),
  \end{array}
	\right.
\end{equation}
and hence the VaR constraint is inactive.

\item $\underline{V} \ge \xi_1, \alpha < \alpha_1(\lambda,\beta_1)$. The maximizer is given by 
\begin{equation}\label{eq:G7,original}
	G _{\lambda ,\underline{V},\alpha}(s):=\left\{ 
	\begin{array}{ll}
K + \frac{1}{n} \left( I\left(\frac{1}{n} \lambda F_{Z} ^{-1}(1- s)\right) - w\right) & ~ \alpha_2(\lambda) \le  s \le 1,\\
\underline{V} & ~ \alpha \le  s < \alpha_2(\lambda),\\
		0 & ~ 0 \le s < \alpha ,
  \end{array}
	\right.
\end{equation}

\item $\underline{V} \ge \xi_1,  \alpha_1(\lambda,\beta_1)\leq \alpha < \alpha_2(\lambda)$.  The maximizer is given by \begin{equation}\label{eq:G8,original}
	G_{\lambda ,\underline{V},\alpha}(s)=\left\{ 
	\begin{array}{ll}
	K + \frac{1}{n} \left( I\left(\frac{1}{n} \lambda F_{Z} ^{-1}(1- s)\right) - w\right) & ~ \alpha_2(\lambda) \le s \le 1 ,\\
	\underline{V} & ~ \alpha \le  s < \alpha_2(\lambda),\\
K + \frac{1}{n} \left( I\left(\frac{1}{n} \lambda F_{Z} ^{-1}(1- s)\right) - w\right) & ~ \alpha_1(\lambda,\beta_1) \le s < \alpha, \\
0 & ~ 0 \le s < \alpha_1(\lambda,\beta_1).
  \end{array}
	\right.
\end{equation}

\item $\underline{V} \ge \xi_1,  \alpha \ge \alpha_2(\lambda)$. The maximizer is given by 
\begin{equation}\label{eq:G9,original}
	G _{\lambda ,\underline{V},\alpha}(s):=\left\{ 
	\begin{array}{ll}
K + \frac{1}{n} \left( I\left(\frac{1}{n} \lambda F_{Z} ^{-1}(1- s)\right) - w\right) & ~ \alpha_1(\lambda,\beta_1) \le  s \le 1,\\
		0 & ~ 0 \le s < \alpha_1(\lambda,\beta_1) ,
  \end{array}
	\right.
\end{equation}
and hence the VaR constraint is inactive.
\end{enumerate}
Therefore, we obtain the optimal solution to \eqref{prob:QF}, $G=G_{\lambda,\underline{V},\alpha}$ and $u=u_\lambda(t)=\lambda\delta Z(t)$.

\noindent \textit{Step 3. Show the existence of the Lagrange multiplier $\lambda$ under different combinations of $\underline{V}$ and $\alpha$.}

Define
\begin{equation*}
\begin{aligned}
f_{\underline{V},\alpha}(\lambda )=& \int _0^1 F_{Z} ^{-1}(1-s)G_{\lambda  ,\underline{V}, \alpha}(s)ds - \E \left[\delta \int _0^T Z (t) u_{\lambda}(t)dt\right] \\
=& \int _0^1 F_{Z} ^{-1}(1-s)G_{\lambda  ,\underline{V}, \alpha}(s)ds - \lambda h(0), 
\end{aligned}
\end{equation*}
where
\begin{equation}\label{eq:h(t)}
	h(t)=\left\{ 
	\begin{array}{ll}
\frac{\delta ^2}{\theta ^2 -2r}(e^{(\theta ^2 -2r)T} - e^{(\theta ^2 -2r)t}) & ~ \theta ^2  \neq 2r ,\\
\delta ^2 (T-t) & ~ \theta ^2  = 2r.
  \end{array}
	\right.
\end{equation}

Note that since $\alpha_1(\lambda,\cdot)$ and $\alpha_2(\lambda)$ are increasing in $\lambda$, and $K+\frac{1}{n}\left(I(\frac{\lambda}{n}F_Z^{-1}(1-s))-w\right)$ is decreasing in $\lambda$, one can easily show that $f_{\underline{V},\alpha}(\lambda)$ is continuous and strictly decreasing on $(0, \infty)$, and hence any $\lambda$ satisfying $f_{\underline{V},\alpha}(\lambda )=V(0)$ is the unique solution to the budget constraint. Cases (c), (f), and (i) are trivial, as the VaR constraint is inactive and $\lambda=\lambda_0$ solves the equation. For the other cases, we calculate $f_{\underline{V},\alpha}(\lambda_0)$ using the corresponding functional form of the optimal quantile function and compare the value with $V(0)$.

For cases (a), (b), (d), and (e), we have
\begin{align*}
f_{\underline{V},\alpha}(\lambda_0)&\ge \int_{\alpha^*}^1 F_Z^{-1}(1-s)\left(K+\frac{1}{n}\left(I\left(\frac{\lambda_0}{n}F_Z^{-1}(1-s)\right)-w\right)\right)ds-\lambda_0h(0)\\
&>\int_{\alpha_1(\lambda_0,\beta_1)}^1 F_Z^{-1}(1-s)\left(K+\frac{1}{n}\left(I\left(\frac{\lambda_0}{n}F_Z^{-1}(1-s)\right)-w\right)\right)ds-\lambda_0h(0)\\
&=V(0),
\end{align*}
where $\alpha^*$ takes value of $\alpha_1(\lambda_0,\beta_2)$ for case (a), $\alpha_2(\lambda_0)$ for case (d), and $\alpha$ for cases (b) and (e). The first inequality holds because $\alpha<\alpha^*$, and the second inequality holds because $\alpha^*<\alpha_1(\lambda_0,\beta_1)$.

For case (g),
\begin{align*}
f_{\underline{V},\alpha}(\lambda_0)>&
\int_{\alpha_2(\lambda_0)}^1 F_Z^{-1}(1-s)\left(K+\frac{1}{n}\left(I\left(\frac{\lambda_0}{n}F_Z^{-1}(1-s)\right)-w\right)\right)ds\\
&+\int_{\alpha_1(\lambda_0,\beta_1)}^{\alpha_2(\lambda_0) }\underline{V}F_Z^{-1}(1-s)ds-\lambda_0h(0)\\
>&\int_{\alpha_1(\lambda_0,\beta_1)}^1 F_Z^{-1}(1-s)\left(K+\frac{1}{n}\left(I\left(\frac{\lambda_0}{n}F_Z^{-1}(1-s)\right)-w\right)\right)ds-\lambda_0h(0)\\
=&V(0),
\end{align*}
where the second inequality holds because
\[
\underline{V}\ge K+\frac{1}{n}\left(I\left(\frac{\lambda_0}{n}F_Z^{-1}(1-s)\right)-w\right)
\]
for $s\in[\alpha_1(\lambda_0,\beta_1),\alpha_2(\lambda_0)]$.

For case (h),
\begin{align*}
f_{\underline{V},\alpha}(\lambda_0)>&
\int_{\alpha_2(\lambda_0)}^1 F_Z^{-1}(1-s)\left(K+\frac{1}{n}\left(I\left(\frac{\lambda_0}{n}F_Z^{-1}(1-s)\right)-w\right)\right)ds\\
&+\int_{\alpha}^{\alpha_2(\lambda_0) }\underline{V}F_Z^{-1}(1-s)ds\\
&+\int_{\alpha_1(\lambda_0,\beta_1)}^\alpha F_Z^{-1}(1-s)\left(K+\frac{1}{n}\left(I\left(\frac{\lambda_0}{n}F_Z^{-1}(1-s)\right)-w\right)\right)ds-\lambda_0h(0)\\
>&\int_{\alpha_1(\lambda_0,\beta_1)}^1 F_Z^{-1}(1-s)\left(K+\frac{1}{n}\left(I\left(\frac{\lambda_0}{n}F_Z^{-1}(1-s)\right)-w\right)\right)ds-\lambda_0h(0)\\
=&V(0),
\end{align*}
where the second inequality holds because
\[
\underline{V}\ge K+\frac{1}{n}\left(I\left(\frac{\lambda_0}{n}F_Z^{-1}(1-s)\right)-w\right)
\]
for $s\in[\alpha,\alpha_2(\lambda_0)]$.

Therefore, we have shown that for the six non-trivial cases, $f_{\underline{V},\alpha}(\lambda_0)>V(0)$. Since $f_{\underline{V},\alpha}(\cdot)$ is decreasing on $(0,\infty)$ and $\lim_{\lambda\nearrow+\infty}f_{\underline{V},\alpha}(\lambda)=-\infty,$ there exists a unique Lagrange multiplier $\lambda\in[\lambda_0,+\infty)$ such that $f_{\underline{V},\alpha}(\lambda)=V(0)$ holds. Therefore, we have $u_\lambda(t)\ge u_0(t)$.

Setting $V_{\text{VaR}}(T)=G_{\lambda,\underline{V},\alpha}(1-F_Z(Z(T)))$, by Lemma \ref{lemma:quantile}, we $V_{\text{VaR}}(T)$ is the optimal terminal firm value.

\noindent\textit{Step 4. Obtain the optimal project choice $v(t)$.}

By the martingale property, we have
\[
 V_{\text{VaR}}(t) = \frac{1}{Z (t) } \E \left[ Z (T) V_{\text{VaR}}(T) -  \delta \int _t^T Z (s) u_{\text{VaR}} (s)ds\bigg|\mathcal{F}_t\right],
\]
which is the optimal firm value at time $t$ in terms of $Z(t)$.
The SDE of $Z(t)$ is given by
\[
dZ(t)=-rZ(t)dt-\theta Z(t) dW(t).
\]
Applying Ito's formula to $b(t, Z(t)),$ we obtain
\begin{equation}\label{eq:SDE_Vvar}
dV_{\text{VaR}}(t)=\left(\dfrac{\partial V_{\text{VaR}}(t)}{\partial t}-\dfrac{\partial V_{\text{VaR}}(t)}{\partial Z(t)}rZ(t)+\dfrac{1}{2}\dfrac{\partial^2 V_{\text{VaR}}(t)}{\partial Z^2(t)}\theta^2Z^2(t)\right)dt-\dfrac{\partial V_{\text{VaR}}(t)}{\partial Z(t)}\theta Z(t)dW(t)    
\end{equation}

 Matching the diffusion term of the SDEs  (1) and \eqref{eq:SDE_Vvar}, we can derive the optimal project choice
\[
v_{\text{VaR}}(t)= -\theta \frac{\partial V_{\text{VaR}(t)}}{\partial Z(t)}Z(t).
\]

\subsection{Proof of Proposition 4.1}\label{appendix_3}
By Proposition 2.1, the firm value at time $t$ is given by
    \begin{align*}
    V(t)=&\frac{1}{Z(t)}\mathbb{E}\left[Z(T)V(T)-\delta^2\lambda\int_t^TZ^2(s)ds | \mathcal{F}_t\right]\\
    =&\frac{1}{Z(t)}\left(\mathbb{E}\left[Z(T)V(T)| \mathcal{F}_t\right]-\delta^2\lambda Z^2(t)\int_t^T\mathbb{E}[e^{-2(r+\frac{1}{2}\theta^2)(s-t)-2(\theta(W(s)-W(t))}]ds \right)\\
    =&\frac{1}{Z(t)}\mathbb{E}\left[Z(T)V(T)| \mathcal{F}_t\right]-Z(t)\lambda e^{-(\theta^2-2r)t}h(t).
    \end{align*}
    where $h(t)$ is given by  \eqref{eq:h(t)}.\\
For the benchmark manager:
\begin{enumerate}
    \item By Proposition 3.2,
    \begin{align*}  
    \mathbb{E}[Z(T)V_0(T)|\mathcal{F}_t]=&\mathbb{E}\left[Z(T) \left(K+\frac{1}{n}\left(I\left(\frac{1}{n}\lambda_0Z(T)\right)-w\right)\right)\times \mathbf{1}_{\{Z(T)\le Z_0\}}|\mathcal{F}_t\right]\\
    =&\left(K-\frac{w}{n}\right)Z(t)e^{-r(T-t)}N(d_2(Z_0)) +n^{\frac{1-\gamma}{\gamma}}\lambda_0^{-\frac{1}{\gamma}}e^{\Gamma(t)}Z(t)^{\frac{\gamma-1}{\gamma}}N(d_1(Z_0)).
    \end{align*}
Rearranging terms, we obtain the expression of $V_0(t)$.

\item By Proposition 3.2, the optimal project choice is 
\begin{equation*}
v_0(t) = - \theta \frac{\partial V_0(t)}{\partial Z(t)}Z(t).
\end{equation*}

We have 
\[
\frac{\partial N(d_1(x))}{\partial Z(t)}=-\frac{\phi(d_1(x))}{\theta Z(t)\sqrt{T-t}},\qquad\frac{\partial N(d_2(x))}{\partial Z(t)}=-\frac{\phi(d_2(x))}{\theta Z(t)\sqrt{T-t}}.
\]
Direct calculation gives
\begin{align*}
    &\frac{\partial V_0(t)}{\partial Z(t)}=-\left(K-\frac{w}{n}\right)e^{-r(T-t)}\frac{\phi(d_2(Z_0))}{\theta Z(t)\sqrt{T-t}}-\frac{n^{\frac{1-\gamma}{\gamma}}e^{\Gamma(t)}}{\gamma\lambda_0^{\frac{1}{\gamma}} Z(t)^{1+\frac{1}{\gamma}}} \left(\frac{1}{\gamma}
N(d_1(Z_0))+\frac{\phi(d_1(Z_0))}{\theta\sqrt{T-t}}\right)\\
&-e^{-(\theta^2-2r)t}\lambda_0h(t).
\end{align*}
Then we can obtain the expression of $v_0(t)$.
\end{enumerate}
For the VaR manager, we present the calculation steps for the case $\underline{V}<K$, and the proofs of the other two cases are similar. Note that one can verify that $K\le \underline{V} <\xi_1$ and $\underline{V}\ge \xi_1$ will result in the same coefficients, and hence the two cases can be combined.
\begin{enumerate}
    \item 
    By Corollary 3.4, we have
    \begin{align*}
        \mathbb{E}[Z(T)V_{\text{VaR}}(T)|\mathcal{F}_t]=&\mathbb{E}[Z(T)V_{\text{VaR}}(T)\mathbf{1}_{\{Z(T)\le \min(Z_{\text{VaR},1},Z_{\text{VaR}}(\beta_2)\}}|\mathcal{F}_t]\\
        &+\mathbb{E}[Z(T)V_{\text{VaR}}(T)\mathbf{1}_{\{Z_{\text{VaR},1}<Z(T)\le Z_{\text{VaR}}(\beta_1)\}}|\mathcal{F}_t]\\
        &+\mathbb{E}[Z(T)V_{\text{VaR}}(T)\mathbf{1}_{\{Z_{\text{VaR}}(\beta_2)<Z(T)\le Z_{\text{VaR},1}\}}|\mathcal{F}_t]
    \end{align*}
    
    By the SDE of the pricing kernel $Z(t)$, we have
    \[
    Z(T)=Z(t)\exp\left\{-\left(r+\frac{1}{2}\theta^2\right)(T-t)-\theta\sqrt{T-t}Z\right\},
    \]
    where $Z$ is a standard normal random variable.
    Plugging in
    \[
    I\left(\frac{1}{n}\lambda_{\text{VaR}} Z(T)\right)=\left(\frac{1}{n}\lambda_{\text{VaR}} Z(T)\right)^{-\frac{1}{\gamma}}
    \]
    and by direct calculation, we obtain
    \begin{fleqn}[0pt]
    \begin{align*}
       &\mathbb{E}[Z(T)V_{\text{VaR}}(T)\mathbf{1}_{\{Z(T)\le \min(Z_{\text{VaR},1},Z_{\text{VaR}}(\beta_2))\}}|\mathcal{F}_t]\\
       =&\left(K-\frac{w}{n}\right)Z(t)e^{-r(T-t)}N(d_2(\min(Z_{\text{VaR},1},Z_{\text{VaR}}(\beta_2))))  \\
       &+n^{\frac{1-\gamma}{\gamma}}\lambda_{\text{VaR}}^{-\frac{1}{\gamma}}e^{\Gamma(t)}Z(t)^{\frac{\gamma-1}{\gamma}}N(d_1(\min(Z_{\text{VaR},1},Z_{\text{VaR}}(\beta_2))),
    \end{align*}
    \begin{align*}
        &\mathbb{E}[Z(T)V_{\text{VaR}}(T)\mathbf{1}_{\{Z_{\text{VaR},1}<Z(T)\le Z_{\text{VaR}}(\beta_1)\}}|\mathcal{F}_t]\\
        =& \left(K-\frac{w}{n}\right)Z(t)e^{-r(T-t)}\left(N(d_2(Z_{\text{VaR}}(\beta_1)))-N(d_2(Z_{\text{VaR},1}))\right)^+\\
        &+n^{\frac{1-\gamma}{\gamma}}\lambda_{\text{VaR}}^{-\frac{1}{\gamma}}e^{\Gamma(t)}Z(t)^{\frac{\gamma-1}{\gamma}}\left( N(d_1(Z_{\text{VaR}}(\beta_1)))-N(d_1(Z_{\text{VaR},1}))\right)^+,
    \end{align*}
    and
    \begin{align*}
        &\mathbb{E}[Z(T)V_{\text{VaR}}(T)\mathbf{1}_{\{Z_{\text{VaR}}(\beta_2)<Z(T)\le Z_{\text{VaR},1}\}}|\mathcal{F}_t]\\
        =&Z(t)\underline{V}e^{-r(T-t)}\left(N(d_2(Z_{\text{VaR},1}))-N(d_2(Z_{\text{VaR}}(\beta_2)))\right)^+.
    \end{align*}
Combining the three terms, the conditional expectation is given by
\begin{align*}
            \mathbb{E}&[Z(T)V_{\text{VaR}}(T)|\mathcal{F}_t]\\
            =&A_1\left(K-\frac{w}{n}\right)Z(t)e^{-r(T-t)}
            +A_2n^{\frac{1-\gamma}{\gamma}}\lambda_{\text{VaR}}^{-\frac{1}{\gamma}}e^{\Gamma(t)}Z(t)^{\frac{\gamma-1}{\gamma}}
            +A_3Z(t)\underline{V}e^{-r(T-t)} .
        \end{align*}
Rearranging terms, we obtain the expression of $V_{\text{VaR}}(t)$.

\item 
By Proposition 3.3, the optimal project choice is
\[
v_{\text{VaR}}(t)=-\theta \frac{\partial V_{\text{VaR}(t)}}{\partial Z(t)}Z(t).
\]
Direct calculation gives 
\begin{align*}
&\frac{\partial V_{\text{VaR}(t)}}{\partial Z(t)}\\
=&-B_1\frac{\left(K-\frac{w}{n}\right)e^{-r(T-t)}}{\theta Z(t)\sqrt{T-t}}-\frac{n^{\frac{1-\gamma}{\gamma}}e^{\Gamma(t)}}{\lambda_{\text{VaR}}^{\frac{1}{\gamma}} Z(t)^{1+\frac{1}{\gamma}}}\left(\frac{1}{\gamma}B_2+\frac{1}{\theta\sqrt{T-t}}B_3\right)-B_4\frac{\underline{V}e^{-r(T-t)}}{\theta Z(t)\sqrt{T-t}}\\
&-e^{-(\theta^2-2r)t}\lambda_{\text{VaR}}h(t).
\end{align*}
Then we can obtain the expression of $v_{\text{VaR}}(t)$.
\end{fleqn}
\end{enumerate}
\end{appendices}

\bibliographystyle{apalike}
\bibliography{ref}

\end{document}